\documentclass[journal=jacsat,manuscript=article]{achemso}

\usepackage[version=3]{mhchem} 
\usepackage{graphicx}
\usepackage{subcaption}
\usepackage[utf8]{inputenc}
\usepackage{hyperref}
\usepackage{array}
\usepackage{tabulary}
\usepackage{svg}
\usepackage{amssymb}
\newcolumntype{s}[1]{>{\centering\arraybackslash}p{3.5cm}}
\usepackage{amsmath}
\usepackage{multirow}   
\usepackage{hyperref}
\usepackage{booktabs}   
\usepackage{tikz}
\usepackage[export]{adjustbox} 
\usepackage[font=small,labelfont=bf]{caption}

\author{Sayeedul I. Sheikh}
\altaffiliation{These authors contributed equally.}
\affiliation[ IITK ]{Department of Chemical Engineering, Indian Institute of Technology Kanpur, Uttar Pradesh 208016, India}

\author{V. Subhasree Navya}
\altaffiliation{These authors contributed equally.}
\affiliation[ IITK ]{Department of Chemical Engineering, Indian Institute of Technology Kanpur, Uttar Pradesh 208016, India}

\author{Riya Sharma}
\altaffiliation{These authors contributed equally.}
\affiliation[ IITK ]{Department of Chemical Engineering, Indian Institute of Technology Kanpur, Uttar Pradesh 208016, India}

\author{Sudip Roy}
\affiliation[ PI ]{Prescience Insilico, Bangalore, India}

\author{Jayant K. Singh}
\affiliation[ IITK ]{Department of Chemical Engineering, Indian Institute of Technology Kanpur, Uttar Pradesh 208016, India}
\alsoaffiliation[ PI ]{Prescience Insilico, Bangalore, India}
\email{jayantks@iitk.ac.in}
\phone{+91 (0)512 2596141}
\fax{+91-512-2590104}

\title[An \textsf{achemso} demo]
  {Digital Surfactant}

\abbreviations{IR,NMR,UV}

\begin{document}

\begin{abstract}
Surfactants play an important role in determining the cleaning performance and stability of detergents. However, the design of new surfactants using traditional methods is often time-consuming, complex, and largely based on trial and error. Recent studies have incorporated data-driven and computational approaches to generate new surfactants and predict properties of surfactants, but most of these approaches either optimize on a single property or train on a small number of surfactants. In this work, we investigate the generative capabilities of an existing graph diffusion based inverse design model and a transformer based molecule optimization model, for non-ionic surfactants. We train both models to generate non-ionic surfactants based on single- and multi-property values, predict the same properties o using trained property predictor models for generated molecules, and validate a few them using molecular dynamics simulations. Our results reveal that the inverse design model is better at generating a diverse set of molecules, while the transformer is better at generating molecules which satisfy input property constraints better. We also observe that molecules generated using single property condition, on average, satisfy the input property condition better when compared to molecules generated using multiple property conditions. From molecular dynamics simulations, we observe that the predicted properties of the selected molecules are close to the simulated results concluding that both methods are capable of generating surfactants that actually satisfy input property condtions.
\end{abstract}
\section{Introduction}

Materials and molecules are at the heart of the petroleum and chemical industries that drive the economy, providing essential products such as fuels, detergents, polymers, pharmaceuticals, and personal care formulations that sustain daily life of which detergents are an indispensable commodity used by consumers worldwide.\cite{adjiman2021process}
The global market for laundry detergents has reached \$ 66.13 billion in 2024 expected to grow to \$ 83.51 billion in 2029 with a compound annual growth rate (CARG) of of 4.9\%.\cite{TBRC_LaundryDetergent_2025}
They are complex mixtures of surfactants, builders, bleaching agents, enzymes and other minor additives\cite{pedrazzani2012biodegradability}. Surfactants account for approximately 37\% of the detergent chemical market share due to their versatility in applications ranging from household cleaning to pharmaceuticals\cite{tanford1978hydrophobic,bussi2007canonical,sugihara2008review,sharma2025morphological}.

The discovery of new surfactants is a complex, time-consuming and resource intensive process. In practice, researchers often begin by identifying existing surfactants which can be ionic, non-ionic, or zwitter-ionic with properties close to the desired performance targets and then modify their molecular structures through iterative experimentation which is largely trial-and-error approach\cite{Abdallah2022Review,Cheng2016Design,Tripathy2023AI,Rosen2012Surfactants}. In recent decades the growing awareness of environmental issues has prompted industry to seek sustainable alternatives for conventional chemical processes and products.\cite{gonzalez2025design} Researchers use machine learning (ML) and active learning (AL) to design the candidate material based on desired performance requirements quickly, which can be tested using computer simulation. This approach can significantly improve the development process and improve efficiency\cite{zuccarini2024material} across multiple sectors, from detergents and home-care products to pharmaceutical and personal-care formulations. Generative Artificial Intelligence is revolutioninzing molecular design by providing efficient tools for generating novel molecular structures tailored to specific functional properties. AI-driven design frameworks can yield high-performance, biodegradable, and energy-efficient surfactants tailored for next-generation sustainable applications\cite{gonzalez2025design,zeng2022deep}.

Previously, for molecular generation and property prediction tasks various method have been proposed. Honda et al.\cite{honda2019smiles} proposed using SMILES\cite{weininger1988smiles} representations of molecules to pretrain transformers\cite{vaswani2017attention} in an unsupervised manner and produce continuous data driven fingerprints of molecules. These fingerprints grasp the semantics of molecules and can be fed to arbitrary predictive models for many downstream tasks. 
Works like Chemformer\cite{irwin2022chemformer} used language models to to pretrain on SMILES representation of a large dataset of molecules, which was further finetuned for various downstream tasks like reaction prediction, molecular optimization and property prediction. SLEFIES representation of molecules have been used in works like MOLGEN\cite{fang2023domain} to pretrain molecular language models on huge corpora of molecules to improve generation quality and latent-space reliability for downstream optimisation and sampling tasks. Text2Mol\cite{edwards-etal-2021-text2mol}, nach0\cite{livne2024nach0} and MolT5\cite{edwards2022translation} use multi modal frameworks to combine text based inputs with other relevant modalities for molecular generation, property prediction, retrieval and synthesis planning. \\
Apart from transformers, score-based generative diffusion models\cite{song2020score} proposed in the field of computer vision, have been extended to unconditional and conditional generation of molecular graphs in Jo et al.\cite{jo2022score} and GDSS\cite{lee2023exploring} respectively, where the graph features were pertubed and subsequently denoised in the continuous space using a proposed system of forward and reverse stochastic differential equations. DiGress\cite{vignac2022digress} proposed a discrete noising method dependant on the marginal distribution of the atoms and bonds, to perturb discrete 2D molecular graph features. Both DiGress\cite{vignac2022digress} and GDSS\cite{lee2023exploring} use predictor models to guide the conditional generation process, while GraphDiT\cite{liu2024graph} couples discrete noise with a predictor free guidance\cite{ho2022classifier} for conditional generation tasks. Diffusion models that generate molecules by leveraging the 3D atomic coordinates while preserving the equivariance properties of molecular graphs have been proposed in works such as Hoogeboom et al.\cite{hoogeboom2022equivariant}, Bao et al.\cite{bao2022equivariant} and Xu et al.\cite{xu2023geometric} Since many surfactant molecules are large, calculation of atom coordinates is expensive and not feasible, hence we focus on methods limited to generation of 1D SMILES and 2D molecular graphs.




 As nonionic surfactants are effective at stabilizing emulsions. The detergency, wetting, and general usefulness of nonionic surfactants make them widely used in industrial and household products, especially detergents.\cite{cheng2020design} For non ionic surfactants, recent works using data-driven methods have shown promise in accelerating surfactant design. 
Nandili et al.\cite{Nnadili2024AIDesign} trained a Variational Autoencoder(VAE)\cite{kingma2013auto} using SLEFIES\cite{krenn2020self} representation of 285 non-ionic surfactants utilizing transfer learning to pre-train the VAE and then reinforcement learning to bias the latent space of the VAE and generate surfactants with lower CMC. Saliency maps were used to interpret important substructures. Martin et al.\cite{gonzalez2025design} used computer-aided molecular design using head-tail combinatorial for multi objective optimization focusing on properties are CMC, Kraft point, Surface Tension, Toxicity, and biodegradability. Recent works have explored the use of large language models (LLMs) as autonomous agents for molecular design. In the context of surfactants, dZiner\cite{ansari2024dziner} employed an LLM to propose new molecules by combining a given input structure with surfactant-specific domain knowledge extracted from the literature. The model was able to generate candidates with improved properties like lower critical micelle concentration (CMC) and target molecules were substantially different from starting molecule as indicated by the low Tanimoto similarity scores. \\
Once new surfactants are generated, their properties must be predicted to evaluate the performance of generative models. Numerous data-driven methods have been employed in recent years for prediction of surfactant properties. Molecular descriptors like connectivity indices and valence connectivity indices were used in\cite{mozrzymas2010prediction}, to predict the critical micelle concentration values of non-ionic surfactants. Qin et al.\cite{qin2021predicting} used Graph Convolutional networks\cite{kipf2016semi} on 2D molecular graph features to predict $\text{CMC}$ values, while machine learning approaches have also been used to predict the phase diagrams of surfactant molecules in\cite{thacker2023can}. More recently, Soares et al.\cite{soares2025representing} used $\text{SMI-TED}_{\text{298M}}$\cite{soares2024large}, a large scale encoder-decoder based foundational model pre-trained on 91 million SMILES samples from PubChem, to predict phase diagrams of non-ionic surfactants. Despite the small dataset of only 60 molecules, the performance of the fine-tuned model on a test dataset showed excellent F1 scores, highlighting the representational capabilities of large pre-trained models for surfactants. SurfPro\cite{hodl2025surfpro} recently introduced a large dataset consisting of 1624 unique surfactant molecules, with a wide variety of experimental property values labelled for each surfactant. The authors also trained AttentiveFP\cite{xiong2019pushing} models to predict single and multiple property values from 2D molecular graphs. \\
While progress has been made into accelerating surfactant design, machine learning methods from other relevant fields have rarely been used for generating surfactants. In addition to this, most previous works use datasets having a smaller number of surfactant molecules\cite{thacker2023can,qin2021predicting}, or focused on using single properties\cite{qin2021predicting, seddon2022machine}. In this work we use two existing molecule generation methods and train them to generate non-ionic surfactant molecules that satisfy input property conditions. Training of generative models is done using the non-ionic surfactants present in the SurfPro dataset, due to its comparatively larger size. To validate the generated molecules, we use property predictor models trained on the entire SurfPro dataset to first screen the generated molecules and then perform Molecular Dynamics simulations on few of the final screened molecules to to evaluate if the models are capable of satisfying input property constraints. Additionally, we also evaluate the performance of models conditioned on a single input property and compare their generative capabilities to models conditioned on multiple properties. The study aims to accelerate the discovery of novel surfactants with tailored properties for industrial applications. 
\section{Methodology}
\subsection{Dataset}
\label{sec:dataset}
The SurfPro\cite{hodl2025surfpro} dataset comprises 1624 surfactant molecules, divided into 8 different surfactant types(Cationic, Gemini cationic, Anionic, Gemini anionic, Non-ionic, Sugar-based non-ionic, Zwitterionic and Gemini zwitterionic). Using experimental results from existing literature, Hödl et al. label some or all of the following property values for each of the molecules in the dataset: Critical micelle concentration(CMC), Air-water surface tension at CMC, Maximum surface excess concentration, Adsorption Efficiency, Surface pressure at CMC and Area at air-water interface. Out of the 1624 molecules, only 647 molecules  have all 6 property values labelled. \\
SurfPro is divided into train and test sets, with the training set consisting of 1484 surfactants randomly divided into 10 folds, while the test dataset consists of 140 surfactant . Using 10-fold cross validation method, AttentiveFP\cite{xiong2019pushing} networks with hidden dimension set to 64, output dimension set to 128 followed by 2 fully connected layers, are trained to predict the earlier stated surfactant properties  for a given graph representation of a molecule. The missing property values of the remaining 977 molecules are then imputed using ensemble averaged predictions of the trained models. \\
In this work, we train generative models using non-ionic and sugar-based non-ionic surfactants, and use their $\text{pCMC}$, $\text{AW\_ST\_CMC}$ and $\text{Area\_min}$ values as the input properties for conditional generation. Out of the 425 molecules of interest(train and test set combined), less than 35\% of the molecules have all three property values experimentally available in the dataset. Due to this constraint, we use model imputed property values computed in Hôdl et al.\cite{hodl2025surfpro}, wherever the actual property values of molecules are missing. 
Finally we train the generative models using 388 non-ionic surfactant molecules and their property values that belong to the original training split of the SurfPro dataset, while the property values of the 37 non-ionic surfactants present in the test split are used as input conditions to generate new molecules from the trained model to evaluate the models' generative capabilities.

\subsection{Inverse Design using Diffusion}
Diffusion models have been used extensively in computer vision tasks\cite{sohl2015deep,ho2020denoising,dhariwal2021diffusion} to generate realistic images and visual content. The forward diffusion process, defined as $q(\mathbf{X}^{1:T}\mid \mathbf{X}^0)=\prod_{t=1}^{T}q(\mathbf{X}^t\mid \mathbf{X}^{t-1})$ gradually corrupts the input datapoint $\mathbf{X}^0$, till it represents a sample from a stationary distrbution $q(\mathbf{X}^T\mid\mathbf{X}^0) = q(\mathbf{X}^T)$ at timestep $t=T\to\infty$. The reverse diffusion process involves learning a parameterized generative model: $p_{\theta}(\mathbf{X}^{0:T})=q(\mathbf{X}^T)\prod_{t=1}^{T}p_{\theta}(\mathbf{X}^{t-1}\mid\mathbf{X}^t, \mathbf{X}^0)$, which is trained to recover the original data from the corrupted states. During inference, new samples are generated by using the trained neural network to iteratively denoise random noise samples from $q(\mathbf{X}^T)$.\\   
In this work, the Graph Diffusion Transformer(GraphDiT)\cite{liu2024graph} has been used to generate new surfactants conditioned on surfactant properties discussed earlier. GraphDiT uses one-hot encoding to encode the atoms of a molecule into node features, $\mathbf{X}_V\in\mathbb{R}^{N\times F_V}$, where $N$ represents the number of atoms in the molecule and $F_V$ represents the total number of possible atom types. A similar approach is followed to encode the bonds as graph edge features, $\mathbf{X}_E\in\mathbb{R}^{N\times N\times F_E}$, where $F_E$ represents the number of possible bond types. Using $\mathbf{X}_V$ and $\mathbf{X}_E$, Liu et al.\cite{liu2024graph} construct a unified feature representation, $\mathbf{X}_G\in\mathbb{R}^{N\times F_G}$, where $F_G = F_V+N\cdot F_E$. This matrix is constructed by first flattening along the last 2 dimensions of $\mathbf{X}_E$ to create a $N\times N \cdot F_E$ shaped matrix and then concatenating this flattened matrix to the node feature matrix $\mathbf{X}_V$, along the last dimension.\\
The forward process involves iteratively sampling from below distribution:
\begin{equation}
    \label{eq: graph_fwd_diff}
    \mathbf{q}(\mathbf{X}_G^t\mid \mathbf{X}_G^{t-1}) = \hat{\text{Cat}}(\mathbf{X}_G^t, \tilde{\mathbf{p}}=\mathbf{X}_G^{t-1}\mathbf{Q}_G^t)
\end{equation}
where the transition matrix, $\mathbf{Q}_G^t\in\mathbb{R}^{F_G\times F_G}$, is defined as: 
\begin{equation}
    \mathbf{Q}_G^t = \begin{bmatrix}
        \mathbf{Q}_V^t & \mathbf{1'}_N\otimes\mathbf{Q}_{VE}^t\\
        \mathbf{1}_N\otimes\mathbf{{Q}}_{EV}^t & \mathbf{1}_{N\times N}\otimes\mathbf{Q}_E^t
    \end{bmatrix}
\end{equation}
where $\mathbf{Q}_V^t\in\mathbb{R}^{F_V\times F_V}, \mathbf{Q}_E^t\in\mathbb{R}^{F_E\times F_E}, \mathbf{Q}_{VE}^t\in\mathbb{R}^{F_V\times F_E}, \mathbf{Q}_{EV}^t\in\mathbb{R}^{F_E\times F_V}$ represent node$\to$node, edge$\to$edge, node$\to$edge and edge$\to$node transition probabilities, $\otimes$ denotes the Kronecker Product and $\mathbf{1}_N\in\mathbb{R}^{N\times1},~\mathbf{1'}_N\in\mathbb{R}^{1\times N},~\mathbf{1}_{N\times N}\in\mathbb{R}^{N\times N}$ denote vectors and matrices with all 1 elements. The above transition matrices are computed separately using the set of marginals $\mathbf{m} = \{\mathbf{m}_V\in\mathbb{R}^{F_V}, \mathbf{m}_E\in\mathbb{R}^{F_E}, \mathbf{m}_{VE}\in\mathbb{R}^{F_V\times F_E}, \mathbf{m}_{EV}\in\mathbb{R}^{F_E\times F_V}\}$. These marginals represent the distribution of atom types, bond types, and the co-occurence of an atom type with a bond type. The transition matrix at time step $t$ is computed as: $\mathbf{Q}^t = \alpha^t\mathbf{I} + (1-\alpha^t)\mathbf{1m'}$. A cosine schedule\cite{nichol2021improved} is followed to define $\alpha^t$. The forward distribution at timestep $t$ can also be conditioned only on the initial state $\mathbf{X}_G^0$ as $\mathbf{q}(\mathbf{X}_G^t\mid \mathbf{X}_G^0) = \hat{\text{Cat}}(\mathbf{X}_G^t, \tilde{\mathbf{p}}=\mathbf{X}_G^{0}\mathbf{\bar{Q}}_G^t)$, where $\bar{\mathbf{Q}}_G^t=\prod_{i=1}^t\mathbf{Q}_G^t$. In Eq.~\ref{eq: graph_fwd_diff}, $\tilde{\mathbf{p}}=\mathbf{X}_G^{t-1}\mathbf{Q}_G^{t}$ represents the unnormalized probability distribution of $\mathbf{X}_G^t$. A modified categorical sampling approach $\hat{\text{Cat}}(\cdot)$ is followed to sample $\mathbf{X}_G^t$. The first $F_V$ columns of $\tilde{\mathbf{p}}$ are normalized and then used to sample $\mathbf{X}_V^t$, while the next $N\cdot F_E$ columns are reshaped and then normalized to sample $\mathbf{X}_E^t$. $\mathbf{X}_G^t$ is then constructed by combining the sampled node and edge features to complete the $\hat{\text{Cat}}(\cdot)$ sampling procedure. \\
Given a set of input conditions, $\mathcal{C}=\{c_1, c_2,\ldots,c_m\}$, during the reverse process, a trained denoising neural network, $f_\theta(G^t, \mathcal{C})$, is used estimate the probability distribution, $p_\theta(\hat{G}^0\mid G^t,\mathcal{C})$. The reverse distribution $p_\theta(G^{t-1}\mid G^t)$ can be computed by using the forward distribution $q(G^t\mid G^{t-1})$. For example, in nodes, given the network predictions $p_{\theta}(\hat{v}\mid G^t) \forall~\hat{v}\in\hat{\mathbf{x}}_v$, the reverse distribution at time step $t$: $p_{\theta}(v^{t-1}\mid G^t)$ can be expressed as the marginalization over the different predicted node types $\hat{v}\in \mathbf{\hat{x}}_v$:
\begin{equation}
\label{eq: reverse_diff}
    p_{\theta}(v^{t-1}\mid G^t) = \sum_{\hat{v}\in \mathbf{\hat{x}}_v}p_{\theta}(v^{t-1}, \hat{v}\mid G^t)= \sum_{\hat{v}\in \mathbf{\hat{x}}_v}q(v^{t-1}|\hat{v})p_{\theta}(\hat{v}|G^t)
\end{equation}
For conditional generation the predictor free guidance method\cite{ho2022classifier} is used. For a given set of input properties, $\mathcal{C}=\{c_1, c_2, \ldots,c_m\}$, predictor free guidance tries to estimate the reverse diffusion step conditioned on the input properties, $\hat{p}_\theta(G^{t-1}\mid G^{t}, \mathcal{C})$, without training a property predictor model by expressing it as a linear combination of following terms: 
\begin{equation}
\label{eq: classifier free}
    \log{\hat{p}_{\theta}(G^{t-1}\mid G^{t}, \mathcal{C})}=s\cdot\log{p_\theta(G^{t-1}\mid G^{t},\mathcal{C})} + (1-s)\cdot\log{p_{\theta}(G^{t-1}\mid G^t)}
\end{equation}
where $s$ is the scale of conditional guidance. The distribution conditioned on $\mathcal{C}$ in Eq.~\ref{eq: classifier free} can be computed using, first the denoising model to predict $p_\theta(\hat{G}^0\mid G^t,\mathcal{C})=f_\theta(G^{t}, \mathcal{C})$ and then using Eq.~\ref{eq: reverse_diff} on the model predictions. The distribution that does not depend on $\mathcal{C}$ in Eq.~\ref{eq: classifier free} is computed in a similar manner, but in this case, the denoising model estimates a distribtution not conditioned on $\mathcal{C}$: $p_\theta(\hat{G}^0\mid G^t)=f_\theta(G^{t-1}, \mathcal{C}=\emptyset)$. During training of $f_\theta(\cdot)$, input properties of a fixed number of randomly chosen training samples are dropped, so that the model can learn to predict $\hat{G}^0$, even when no input conditions are given. \\
In the denoising model, the conditions, $c_i\in\mathcal{C}$, are encoded into a $D$-dimensional vector, $\mathbf{c}$, where categorical conditions are one-hot encoded, while regression based conditions are are assigned to learnable cluster centroids. 
At timestep $t$, the noisy graph $G^t$ is first encoded into a hidden space $\mathbf{H} = \text{Linear}(\mathbf{X}_G^t)$ where $\mathbf{H}\in\mathbb{R}^{N\times D}$. Standard Transformer\cite{vaswani2017attention} layers of self attention and multi layer perceptrons are used followed by Adaptive Layer Normalization(AdaLN)\cite{peebles2023scalable}: $\mathbf{H}_{\text{out}}=\text{AdaLN}(\mathbf{H,c})$. For each row $\mathbf{h}$ in $\mathbf{H}$:
\begin{equation}
    \label{eq: adaln}
    \text{AdaLN}(\mathbf{h,c}) = \gamma_{\theta}(\mathbf{c})\odot\frac{\mathbf{h}-\mu(\mathbf{h})}{\sigma(\mathbf{h})}+\beta_{\theta}(\mathbf{c})
\end{equation}
where $\mu(\cdot)$ and $\sigma(\cdot)$ are mean and standard deviation of row, $\odot$ represents element-wise products, and $\gamma_{\theta}(\cdot)$ and $\beta_{\theta}(\cdot)$ are neural networks each consisting of 2 linear layers with SiLU\cite{elfwing2018sigmoid} activation in between the layers. At the final layer $\hat{\mathbf{X}}_G^0$ is predicted using the hidden state $\mathbf{H}$ and multi layer perceptrons($\text{MLP}$) as: $\hat{\mathbf{X}}_G^0 = \text{AdaLN}(\text{MLP}(\mathbf{H}),\mathbf{c})$. The node and edge representations $\hat{\mathbf{X}}_V^0$ and $\hat{\mathbf{X}}_E^0$ are recovered. During training, $t$ is randomly sampled, and the denoiser is trained to accurately reproduce $\mathbf{X}_V^0$ and $\mathbf{X}_E^0$ given the set of conditions $\mathcal{C}$ and noisy graph feature $\mathbf{X}_G^t$. During inference, starting from $t=T$ to $t=1$, given $\hat{\mathbf{X}}_G^t$, probability distributions of $\hat{\mathbf{X}}_V^0$ and $\hat{\mathbf{X}}_E^0$ are predicted using the denoising model for $\mathcal{C}=\mathcal{C}_{\text{desired}}$ and $\mathcal{C}=\emptyset$ . Then Eqs.~\ref{eq: reverse_diff} and Eq.~\ref{eq: classifier free} are used to sample the noisy node and edge features at timestep $t-1$. Once $\hat{p}_{\theta}(G^0\mid G^1,\mathcal{C})$ is estimated, it is used to sample the final generated graph features: $\tilde{\mathbf{X}}_V^0$ and $\tilde{\mathbf{X}}_E^0$. The final molecule constructed by connecting randomly selected atoms, based on the distribution of atom numbers in the training dataset. \\\\
We train the GraphDiT model for 10000 epochs, using the $\text{torch-molecule}$\cite{liu_torchmol} library. The training dataset consisted of the non-ionic molecules present in the training split of the SurfPro dataset. We try to generate 10 molecules for each set of input conditions during sampling. The input conditions are same as the property values of non-ionic surfactant molecules present in the test split of the dataset. Two separate diffusion models are trained, the first model, conditioned only of the CMC values, while the second model is conditioned on the CMC, Air-Water Surface tension at CMC and Area at air water interface values. While training we reduce batch size to 40, and use defaults values for all remaining hyperparameters. $\text{torch-molecule}$ automatically checks if the generated molecule is valid or not during sampling. Resampling, incase an invalid molecule is generated for a given set of input condition, is not done in this work, as this lets us estimate the models' capability to generate valid molecules without assistance.

\subsection{Molecule Optimization using Transformers}

Molecular optimization is analogous to the machine translation problem in natural language processing, where the source molecule, is modified to obtain a generated molecule with desired property modifications , using its SMILES \cite{weininger1988smiles}, a string-based representation. For this, matched molecular pairs (MMPs), a pair of molecules that differ by a single, well-defined structural transformation of surfactant were made using the mmpdb tool\cite{dalke2018mmpdb}, to  guide the Transformer-based generative model.



For this model, we focus on utilizing chemical transformations (i.e. MMPs) to optimize a starting molecule to a promising one. In particular, given a starting molecule and the desirable property changes, the goal is to generate molecules that have desirable properties while being structurally similar to the the starting molecule.



As proposed in He et al.\cite{He2021MolecularOptimization}, given a set of MMPs $\{(X, Y, Z)\}$ where $X$ represents source molecule, $Y$ represents target molecule, and $Z$ represents the property change between source molecule $X$ and target molecule $Y$, the Transformer learns the mapping $(X, Z) \in \mathcal{X} \times \mathcal{Z} \rightarrow Y \in \mathcal{Y}$ during training where $\mathcal{X} \times \mathcal{Z}$ represents the input space and $\mathcal{Y}$ represents the target space. During testing, given a new $(X, Z) \in \mathcal{X} \times \mathcal{Z}$, the models will be expected to generate a diverse set of target molecules with desirable properties.


\textbf{Transformer Architecture}: In this paper, the encoder-decoder Transformer architecture used in He et al.\cite{He2021MolecularOptimization} is explored for surfactant optimization. To construct a vocabulary, all the source and target SMILES in our dataset were tokenized, containing all the possible tokens. The start and end symbols were added as tokens to represent the start and end of a sequence, respectively. To guide the model in generating molecules with specified property constraints, the property changes between source and target molecules were concatenated with the source SMILES strings. Therefore, each possible single property change was treated as a token (e.g. $\Delta$pCMC$\in$(0.3,0.5) and added to the vocabulary. First step is to convert the list of tokens in the input sequence into an embedding vector, followed by a positional encoding, which provides the information about the order of element in a sequence. The resulting modified vector encodes both semantic meaning and position within the sequence.  
The transformer is built around two key modules: the encoder and decoder. The encoder consists of a stack of 6 identical encoder layers. Each token is represented as a vector that passes through these layers. The input for the first encoder layer is the embedding of each token in the input sequence. Each encoder layer has two sublayers: The first is a multi-head self-attention mechanism, and the second is a simple, position-wise fully connected feed-forward network. After that, a residual connection around each of the sub layers, followed by layer normalization. \cite{vaswani2017attention} 
In the first encoder layer, for each positional token embedding at position t in the input sequence is processed through a multi-head self-attention sub-layer is used, which helps the encoder to identify the relevant token in the sequence. The input embedding is linearly projected into three vectors Q (query), K (key), and V (value). These are used to compute the self-attention score that determines the importance of all the tokens in the input sentence for encoding the input token being processed.

\begin{equation}
\label{eq:attention}
\text{Attention}(\mathbf{Q}, \mathbf{K}, \mathbf{V})
= \text{softmax}\!\left( \frac{\mathbf{QK}^\top}{\sqrt{d_k}} \right)\mathbf{V}
\end{equation}

The output is computed as a weighted sum of the values, where the weights assigned to each value are computed by the dot product of the query with all the keys, dividing each by $\sqrt{d_k}$, and applying a softmax function over it, where $d_k$ is the dimensionality of the Key vector. This is the process for single-head attention.

Multi-head attention learns multiple sets of Q, K, V representations, which are later used to obtain attention vectors in parallel, yielding $ d_v$-dimensional output values, where $d_v$ is the dimensionality of the Value vector. These values are then processed by the feed-forward neural network sub-layer. The output of the feed-forward neural network sub-layer is sent to the next encoder layer. The decoder is also made of a stack of N identical decoder layers. Each decoder layer is made up of three sub-layers: masked multi-head self-attention sub-layer, fully connected feed-forward network sub-layer, and encoder-decoder multi-head attention sub-layer. The decoder does the same thing as the encoder, but the attention in the decoder is different from the attention in the encoder in the following ways: (i) the self-attention in the encoder allows for each position attending to all positions from the previous encoder layer whereas the self-attention in the decoder only allows for each position attending to the past positions by masking the future ones. (ii) Moreover, the attention for the encoder-decoder multi-head was added to permit the decoder to accommodate the input sequence better. The uppermost encoder layer’s output is specifically converted into the K and V vectors that are the ones used by every encoder-decoder multi-head attention sub-layer.

The model was trained on a NVIDIA GeForce RTX 4080 using the same hyperparameters as He et al.\cite{He2021MolecularOptimization}


\textbf{Data Preparation for Pre-Training and Finetuning:} The models are pre-trained on a set of MMPs extracted from ChEMBL together with the property changes between the source and target molecules.
Around 198,560 matched molecular pairs (including reverse transformations) without property constraints from He et al \cite{He2021MolecularOptimization} are used to train the model to learn the structural-property relationship. The model is fine-tuned on a set of 1267 MMPs extracted using the mmpdb\cite{dalke2018mmpdb} tool, from the training split of SurfPro dataset, with the property changes between the source and target molecules.


\textbf{Test set: }Each test sample $(X, Z)$ consists of two parts, the starting molecule($X$) which is to be optimized and the desirable property change $Z$, which therefore determines the input data space $X$. In order to evaluate our models we construct a test set by forming MMPs using the test split of the SurfPro\cite{hodl2025surfpro} dataset, forming 22 pairs, where the desirable property changes are determined by the MMPs. Since the starting and target molecules, $X$ and $Y$ belong the test set, these pairs are not seen by the models during training, and these pairs are used to evaluate the models generalizibility.



\subsection{Property Predictor Models}

Following Hödl et al.\cite{hodl2025surfpro}, we train single and multi property predictor GNN models on the entire SurfPro training dataset, to predict the properties of generated molecules. The single property predictor model is an AttentiveFP\cite{xiong2019pushing} network followed by a normalizing layer\cite{ba2016layer} and three linear layers with ReLU activation between them. This model is trained to predict the $\text{pCMC}$ of the input surfactant molecule. The multi property predictor model follows a similar architecture to the single property predictor model, but the last linear layer is changed to predict $\text{pCMC}$, $\text{AW\_ST\_CMC}$ and $\text{Area\_min}$ values simultaneously. The input feature set is kept similar to the one used in Hödl et al.\cite{hodl2025surfpro} and we set the $\textit{hidden channel}$, $\textit{output channel}$, $\textit{num\_layers}$, $\textit{dropout}$ and $\textit{num\_timesteps}$ hyperparameters of AttentiveFP network to $64,~128,~2, ~0.1$ and $2$ respectively, as this model size is found to yield consistent results across all experiments in\cite{hodl2025surfpro} while also having only $116$K learnable parameters. Similarly we used 10 fold cross validation during training, train the model for a maximum of 500 epochs, and implement early stopping after the first 50 epochs based on the dynamics of the mean absolute error(MAE) of the validation set and also find the optimum learning rate. In order to implement early stopping and finding optimum learning rate, PyTorch Lightning's\cite{Falcon_PyTorch_Lightning_2019} $\text{EarlyStopping}$ and $\text{LearningRateFinder}$ methods are used. 
The model parameters were optimized using the AdamW\cite{loshchilov2017decoupled} optimizer and trained with Huber loss at the learning rates proposed by the $\text{LearningRateFinder}$ for each fold.
\subsection{Density Functional Theory (DFT) and Molecular Dynamics (MD) Simulations Details }
In this work, Geometry optimizations were carried out using Density Functional Theory (DFT) with the B3LYP hybrid functional, which combines Becke’s three-parameter exchange functional with the Lee–Yang–Parr correlation functional. For H, C, N, and O atoms, the 6-311++G* basis set was employed, a triple-$\zeta$ split-valence set augmented with diffuse and polarization functions to accurately describe electron density, lone pairs, and weak interactions.[26] Unless otherwise specified, all optimizations were conducted in the gas phase under standard convergence criteria. Frequency calculations were performed at the same level of theory to verify that optimized structures correspond to true minima and to obtain thermodynamic corrections.
For MD simulations, we used GROMACS\cite{abraham2015gromacs,hess2008gromacs} package to perform the all-atom molecular dynamics simulations. The simulation system of the air–water interface, represented as the water-vacuum interface, was designed to mimic the experimental surface coverage of surfactant molecules in line with previous works . Packmol software ~\cite{martinez2009packmol} was used generate the initial configurations with the 8 * 8 * $L_z$ nm$^3$ simulation box, which consists of a $L_z$ nm thick water layer in z-direction, with the surfactant monolayer placed on both sides of the water slab along the XY plane, which covers the 8 * 8 nm$^2$ surface area of the water slab. Solvent was added to maintian desired concentration the configuration was edited to maintain a vacuum of 3 nm  on the upper/lower sides of the water layer to create a vapour-liquid interface. The periodic boundary condition was implemented in all directions of the simulation box. 

The number of surfactant molecules placed in the monolayer was computed by dividing the cross-sectional area of the simulation box by the area per surfactant molecule at desired concentration. The $Area\_{min}$ of surfactant molecule was predicted form the GNN model.

The system was then energy minimized using steepest descent algorithm. In addition, equilibration simulations were performed in the ensemble $NVT$ for 20 ns using the temperature velocity rescale thermostat\cite{swope1982computer}with a coupling constant of 0.1 ps. The final system was then simulated for a production run in a $NVT$ ensemble for 100 ns  wherein the temperature velocity-rescale thermostat  with relaxation time of 0.1 ps. A timestep of 2 fs is used for all the simulations. The long range Coulombic interactions were computed using the Particle-Mesh-Ewald (PME) method with a cut-off of 1.20 nm and a fourier grid spacing of 0.16 nm. Similarly, the van der Waals interactions were shifted and smoothly switched to zero from 1.00 nm to 1.20 nm. Additionally, long range dispersion corrections were added to energy and pressure calculations. The h-bonds were constrained using LINCS algorithm and the coordinates and energies were stored in the log files after every 40 ps.

\section{Results and Discussion}

\subsection{Property Predictor Models}
Following Hödl et al.\cite{hodl2025surfpro}, the final prediction ($y_{\text{pred}}$) is obtained by averaging the outputs of the 10 models trained in the 10-fold cross-validation setup. To evaluate performance, we compute the MAE, Root Mean Squared Error (RMSE), and Coefficient of Determination ($R^2$) scores by comparing the model predictions ($y_{\text{pred}}$) with the target values ($y_{\text{target}}$) for surfactants in the test set. The equations used to compute each metrics are: 
\begin{equation}
    \text{MAE} = \frac{1}{N}\sum_{i=1}^N\mid y_{\text{target}}-y_{\text{pred}}\mid
    \label{eq: mae}
\end{equation}
\begin{equation}
    \text{RMSE} = \sqrt{\frac{1}{N}\sum_{i=1}^N(y_{\text{target}}-y_{\text{pred}})^2}
    \label{eq: rmse}
\end{equation}
\begin{equation}
    \text{R}^2 = 1-\frac{\sum_{i=1}^N(y_{\text{target}}-y_{\text{pred}})^2}{\sum_{i=1}^N(y_{\text{target}}-y_{\text{mean}})^2}
    \label{eq: r2}
\end{equation}

where $N$ is the total number of samples in the test dataset, and $y_{\text{mean}}=\frac{1}{N}\sum_{i=1}^Ny_{\text{target}}$. For models that predict multiple properties, Equations~\ref{eq: mae},\ref{eq: rmse}, and~\ref{eq: r2} are computed separately for each property. 


\begin{table}[h]\centering
\scalebox{0.9}{\begin{tabular}{@{}c|c|c|c|c@{}}
\toprule
\textbf{MODEL TYPE}&\textbf{METRIC}&\textbf{pCMC}&\textbf{AW\_ST\_CMC}&\textbf{Area\_min}\\\midrule
\multirow{3}{*}{\textbf{Single}}&\textbf{MAE}&0.256&-&-\\
&\textbf{RMSE}&0.393&-&-\\
&\textbf{R$^2$}&0.865&-&-\\\midrule
\multirow{3}{*}{\textbf{Multi}}&\textbf{MAE}&0.251&2.514&0.209\\
&\textbf{RMSE}&0.372&3.508&0.322\\
&\textbf{R$^2$}&0.879&0.810&0.764\\\bottomrule
\end{tabular}}
\caption{Results of property predictor models on the test split of SurfPro\cite{hodl2025surfpro} Dataset. In the first column, model type \textbf{Single} refers to the models that were trained to only predict $\text{pCMC}$ values, while \textbf{Multi} refers to the models that predict $\text{pCMC, AW\_ST\_CMC, Area\_min}$ values simultaneously.}
\label{tab:prop_pred}
\end{table}

Table~\ref{tab:prop_pred} shows the performance of the property predictor models on the test split of SurfPro dataset. Our results are close to the results reported by Hödl et al.\cite{hodl2025surfpro} for same model size. The \textbf{Single} model has been employed later to predict $\text{pCMC}$ values of molecules generated from generative model conditioned on $\text{pCMC}$ values only, while the \textbf{Multi} model has been used to predict all three property values for molecules generated from the generative model conditioned on $\text{pCMC, AW\_ST\_CMC}$, and $\text{Area\_min}$ values.

\subsection{Generative Models}

In order to assess the generative model's capability to generate diverse new molecules and ability to cover the entire training distribution, we follow Liu et al.\cite{liu2024graph} to compute the following metrics\cite{polykovskiy2020molecular} on the generated molecules: (1) \% of valid generated molecules(Validity), (2) internal diversity of generated molecules(Diversity), which tries to check if model generates a wide variety of samples or suffers from mode collapse by aggregating and normalizing the tanimoto similarity scores between all generated molecules and subtracting this score from 1,
(3) fragment based similarity\cite{degen2008art} with training set(Similarity), which tries to quantify if molecules of generated set have similar fragments to the training set, (4) Fréchet ChemNet Distance(FCD)\cite{preuer2018frechet}, which measures how similar the distribution of generated set is to the training set distribution(Distance), (5) Ratio of number of heavy atoms present in the generated set to the original training set(Coverage), (6) The mean absolute error(MAE) scores of each property value that is predicted using the trained property predictor models discussed earlier. \\
\begin{table}[h]
\centering
\scalebox{0.9}{
\begin{tabular}{@{}c|c|c|c|c|c|c|c|c@{}}
\toprule
\textbf{METHOD}&\textbf{Valid(\%)}($\uparrow$)&\textbf{Div}($\uparrow$)&\textbf{Sim}($\uparrow$)&\textbf{Dist}($\downarrow$)&\textbf{Cov}($\uparrow$)&\textbf{pCMC}($\downarrow$)&\textbf{ST}($\downarrow$)&\textbf{Area}($\downarrow$)\\\midrule
Diff-Single&96.757&0.836&0.962&6.457&7/7&0.975&-&-\\
Trfm-Single&96.429&0.661&0.993&7.401&4/7&0.482&-&-\\
Diff-Multi&96.757&0.840&0.959&5.260&7/7&1.306&4.039&0.286\\
Trfm-Multi&100&0.640&0.996&7.662&4/7&0.695&2.938&0.150\\
\bottomrule
\end{tabular}}
\caption{Metrics to assess the generative capabilities of transformer (Trfm) and diffusion (Diff) models conditioned on different sets of properties. The models conditioned on only $\text{pCMC}$ are referred to as $\text{Single}$, while the mutli-conditioned models are referred to as $\text{Multi}$. Columns 2--6 show the Validity, Diversity, Similarity, Distance and Coverage scores respectively, while columns 7--9 show the MAE values between predicted and target $\text{pCMC}$, $\text{AW\_ST\_CMC}$ and $\text{Area\_min}$ values of the generated set.}
\label{tab:metrics}
\end{table}
For each method, we generate 10 molecules for a given set of properties in the test set. Table~\ref{tab:metrics} shows the earlier discussed metrics computed using all generated molecules for a method. Although the coverage scores of diffusion methods appear higher compared to transformer methods, it must be noted that the transformers are trained to generate molecules that are structurally close to the input surfactant. This leads to lower coverage scores because the input reference molecules in the test set used by the transformer methods do not contain all the unique heavy atoms present in the training set. \\
\begin{figure}[t]
    \centering
    \begin{subfigure}[b]{0.45\textwidth}
        \includegraphics[width=\textwidth]{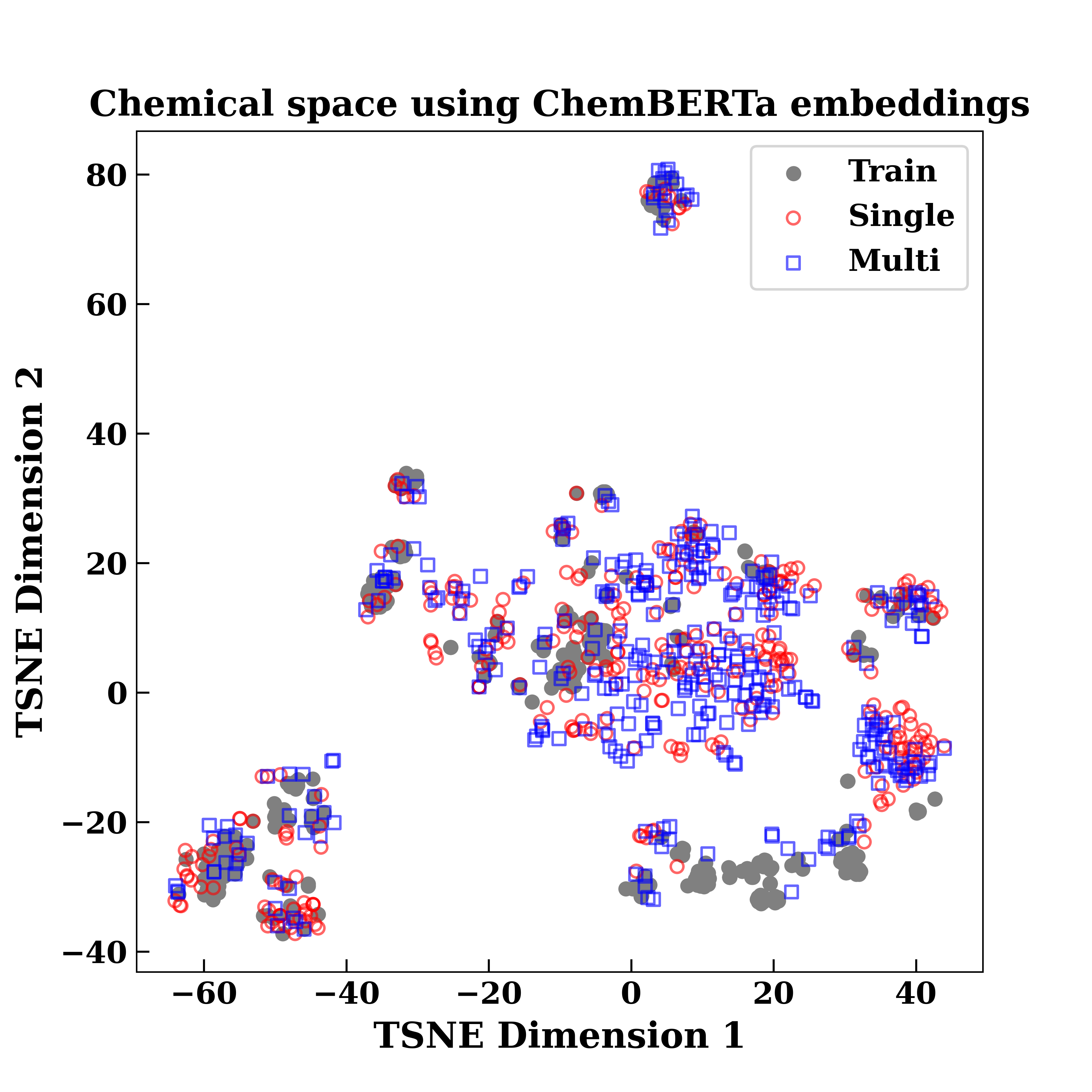}
        \caption{tSNE applied to ChemBERTa embeddings of molecules generated using Diffusion models.}
        \label{fig:tsne(a)}
    \end{subfigure}
    \hfill
    \begin{subfigure}[b]{0.45\textwidth}
        \includegraphics[width=\textwidth]{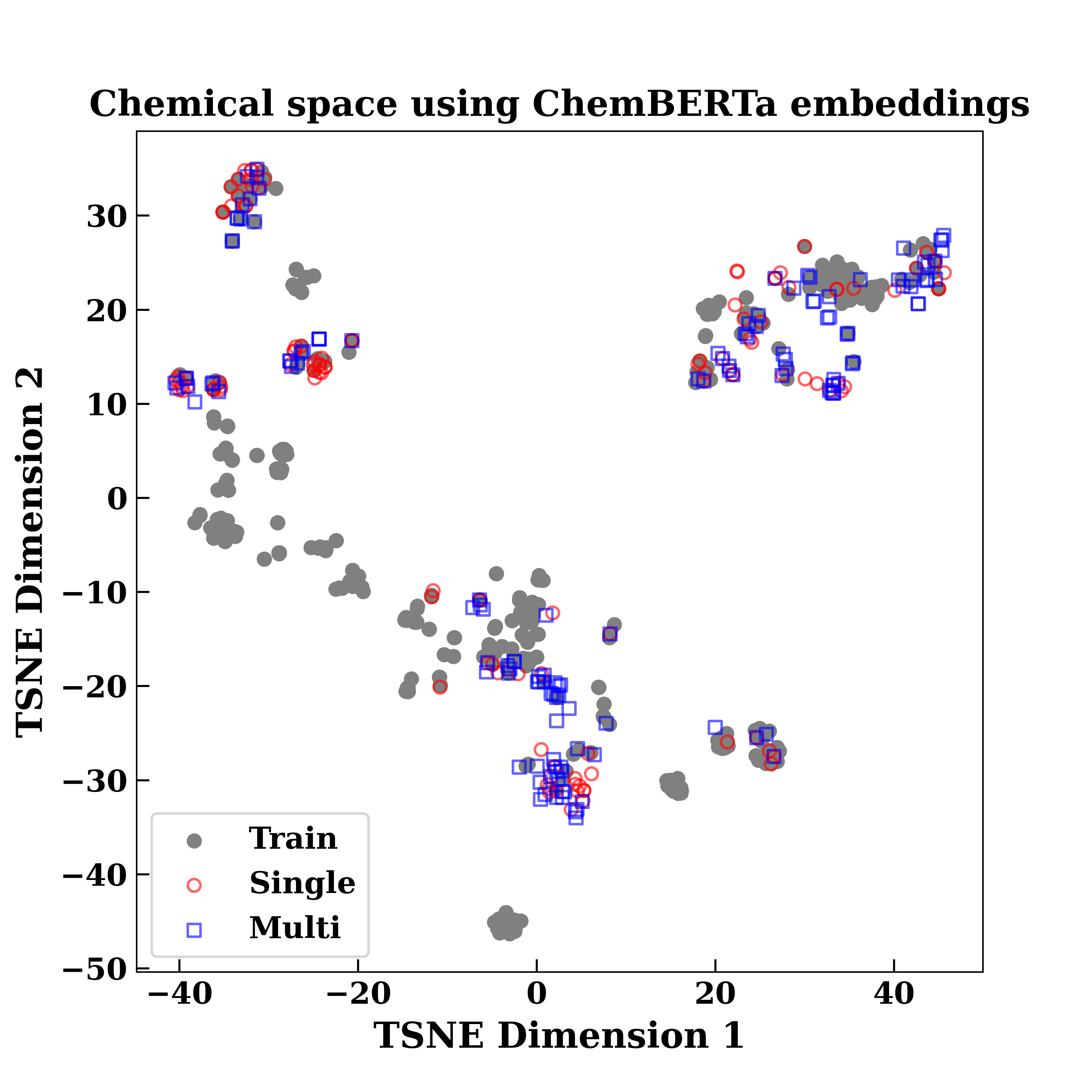}
        \caption{tSNE applied to ChemBERTa embeddings of molecules generated using Transformer models.}
        \label{fig:tsne(b)}
    \end{subfigure}
    \caption{Visualizing and comparing the distribution of generated molecules with the original training set by reducing the output dimensions of ChemBERTa embeddings. Figure~\ref{fig:tsne(a)} shows the space of generated molecules and training set using diffusion while Figure~\ref{fig:tsne(b)} shows the same but the molecules are generated using transformer models. The molecules that are generated using only $\text{pCMC}$ values are shown in red in the both figures while molecules generated using multi-property constraints are shown in blue. Grey is used to visualize the embeddings of the molecules that belong to the trainign set.}
    \label{fig:tsne}
\end{figure}
From Table~\ref{tab:metrics}, we can see that diffusion methods generate a more diverse set of molecules compared to transformers, showing a 26\% improvement, partially because transformers were trained to generate molecules that changed the input molecules minimally, while satisfying property constraints. This causes the transformer methods to acheive higher similarity scores when compared to the fragments of the training dataset. However the distribution of molecules generated using diffusion methods is comparatively closer to the training dataset, when compared to the transformer approaches, which is also seen in Figure~\ref{fig:tsne}, where we use a pre-trained ChemBERTa\cite{chithrananda2020chemberta} model to to get output representations of training and generated molecules and then perform tSNE\cite{maaten2008visualizing} on the output embeddings to visualize the training and generated molecule space. Although the transformer generates molecules whose fragments resemble the training dataset better, the distance scores suggest that the diffusion methods are better at generating set of molecules that resemble the training dataset distribution better. In Figure~\ref{fig:tsne(a)} we can see that molecules generated using diffusion almost occupy the entire space occupied by training molecules, while in Figure~\ref{fig:tsne(b)} we see that the molecules generated using the transformer approaches only occupy certain parts of the training space. However, the molecules generated by the transformers produce significantly lower MAE values when compared to target property values, which suggests that transformers can generate molecules that satisfy property values closely, while diffusion models are able to generate a diverse set of molecules. 
These results suggest that 
\\ 
\begin{figure}[t]
    \centering
    \begin{subfigure}[b]{0.45\textwidth}
        \includegraphics[width=\textwidth]{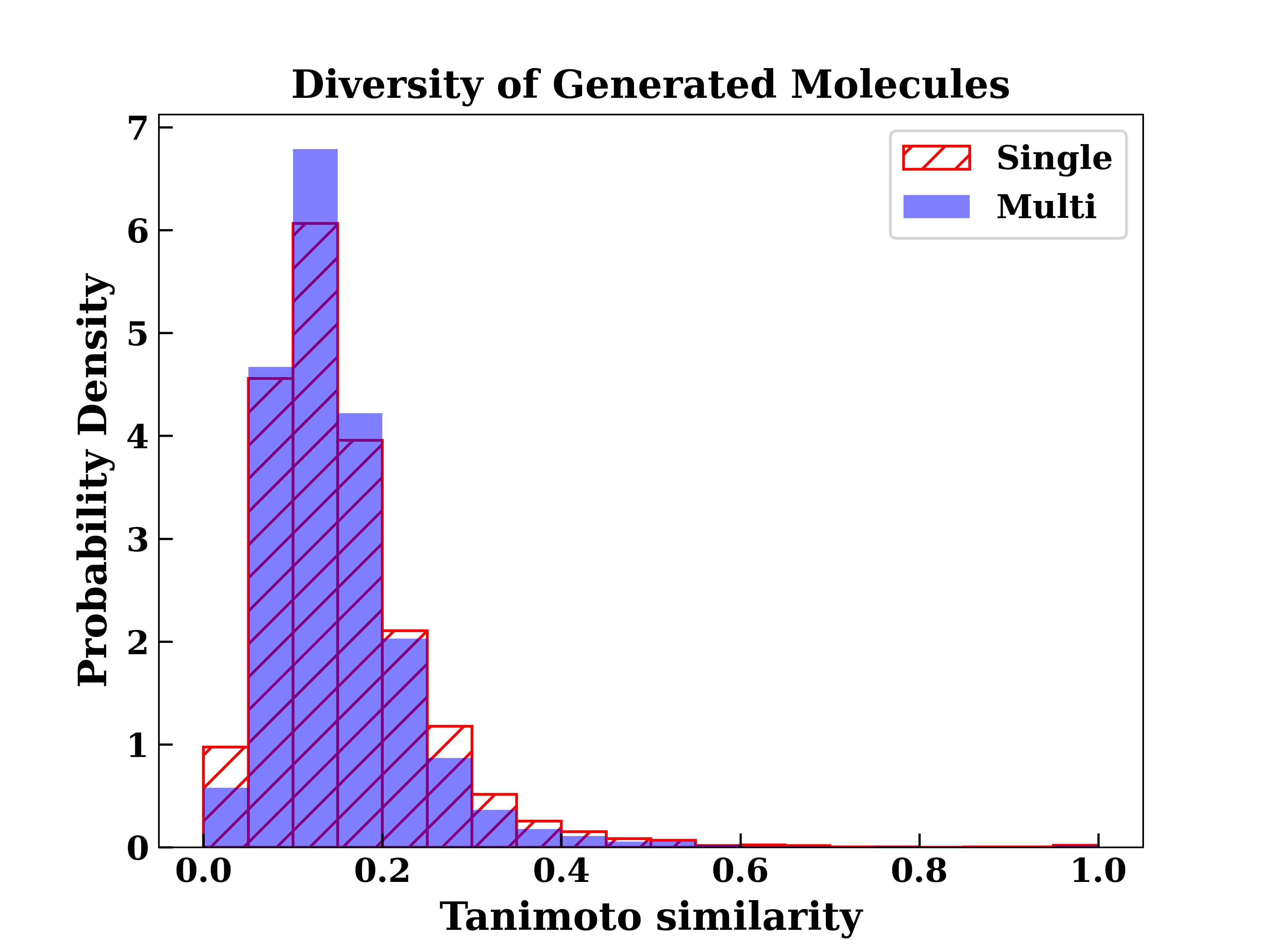}
        \caption{Diffusion}
        \label{fig:tanimoto(a)}
    \end{subfigure}
    \hfill
    \begin{subfigure}[b]{0.45\textwidth}
        \includegraphics[width=\textwidth]{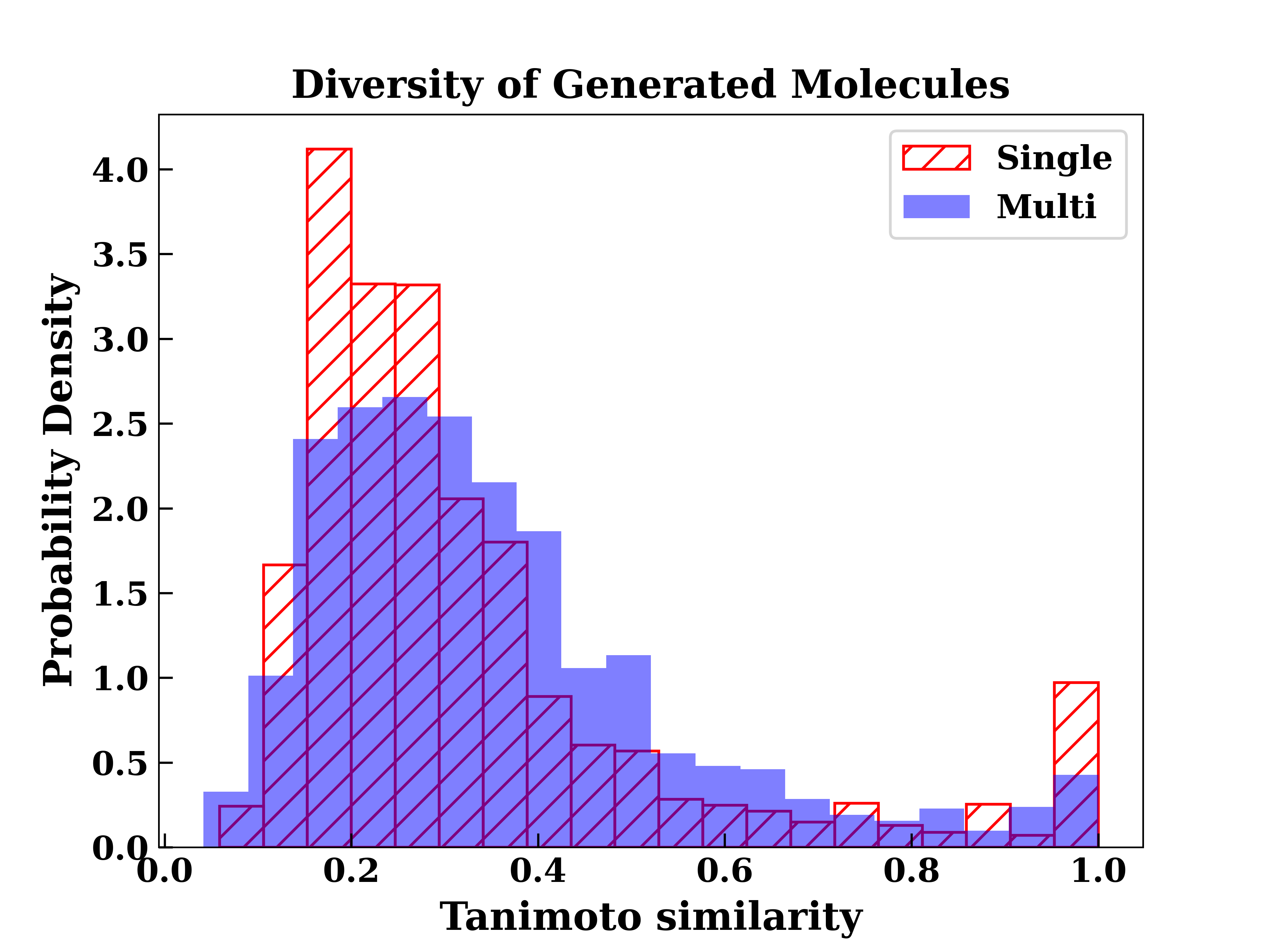}
        \caption{Transformer}
        \label{fig:tanimoto(b)}
    \end{subfigure}
    \caption{Morgan fingerprints were computed for each generated molecule using RDKit's \textit{GetMorganFingerprintasBitVec} function setting radius to 2 and nbits to 2048. These fingerprints were used to compute the tanimoto similarity scores between each of the generated molecules for a particular method. The distribution of these scores is shown in the above figures with red indicating the distribution of scores for molecules generated using only on $\text{pCMC}$ values while blue shows the distribution of scores molecules generated using all three property values.}
    \label{fig:tanimoto}
\end{figure}
Comparing multi to single methods in Table~\ref{tab:metrics}, for Diffusion models we observe an 18\% decrease in distance score for multi-property conditioned models while the validity, diversity, similarity, and coverage scores  show negligible variation. In transformers, negligible variation is observed in validity, diversity, similarity, distance and coverage scores for the two methods. However, in both Transformer and Diffusion approaches, we observe that the models conditioned only on $\text{pCMC}$ generate molecules that satisfy the target $\text{pCMC}$ values better than multi conditioned models.

In Figure~\ref{fig:tanimoto}, we plot the distribution of tanimoto similarity scores computed between all generated valid molecules of a particular method, to compare if a particular method is capable for producing more diverse set of molecules. From Figure~\ref{fig:tanimoto(a)}, we conclude that in diffusion both single and multi conditioned models can generate a diverse set of molecules through diffusion, with molecules generated using only $\text{pCMC}$ values being as diverse as the molecules generated from model conditioned on all 3 property values. For transformers, we observe more varied results in Figure~\ref{fig:tanimoto(b)}, where the model conditioned on $\text{pCMC}$ is capable of generating a more diverse set of molecules compared to the multi conditioned transformer model. 

\begin{figure}[t]
    \centering
    \begin{subfigure}[b]{0.45\textwidth}
        \includegraphics[width=\textwidth]{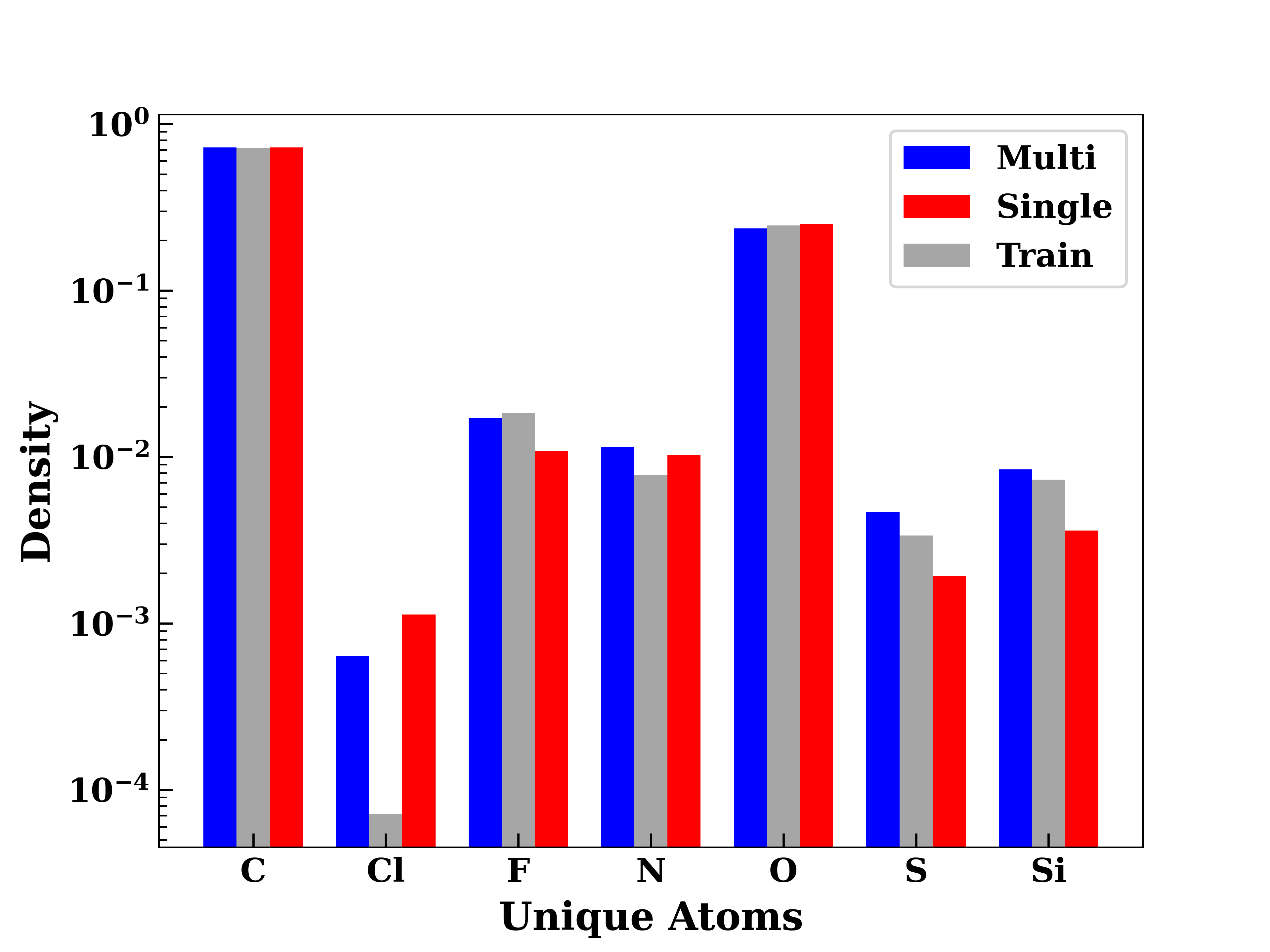}
        \caption{Diffusion}
        \label{fig:atoms(a)}
    \end{subfigure}
    \hfill
    \begin{subfigure}[b]{0.45\textwidth}
        \includegraphics[width=\textwidth]{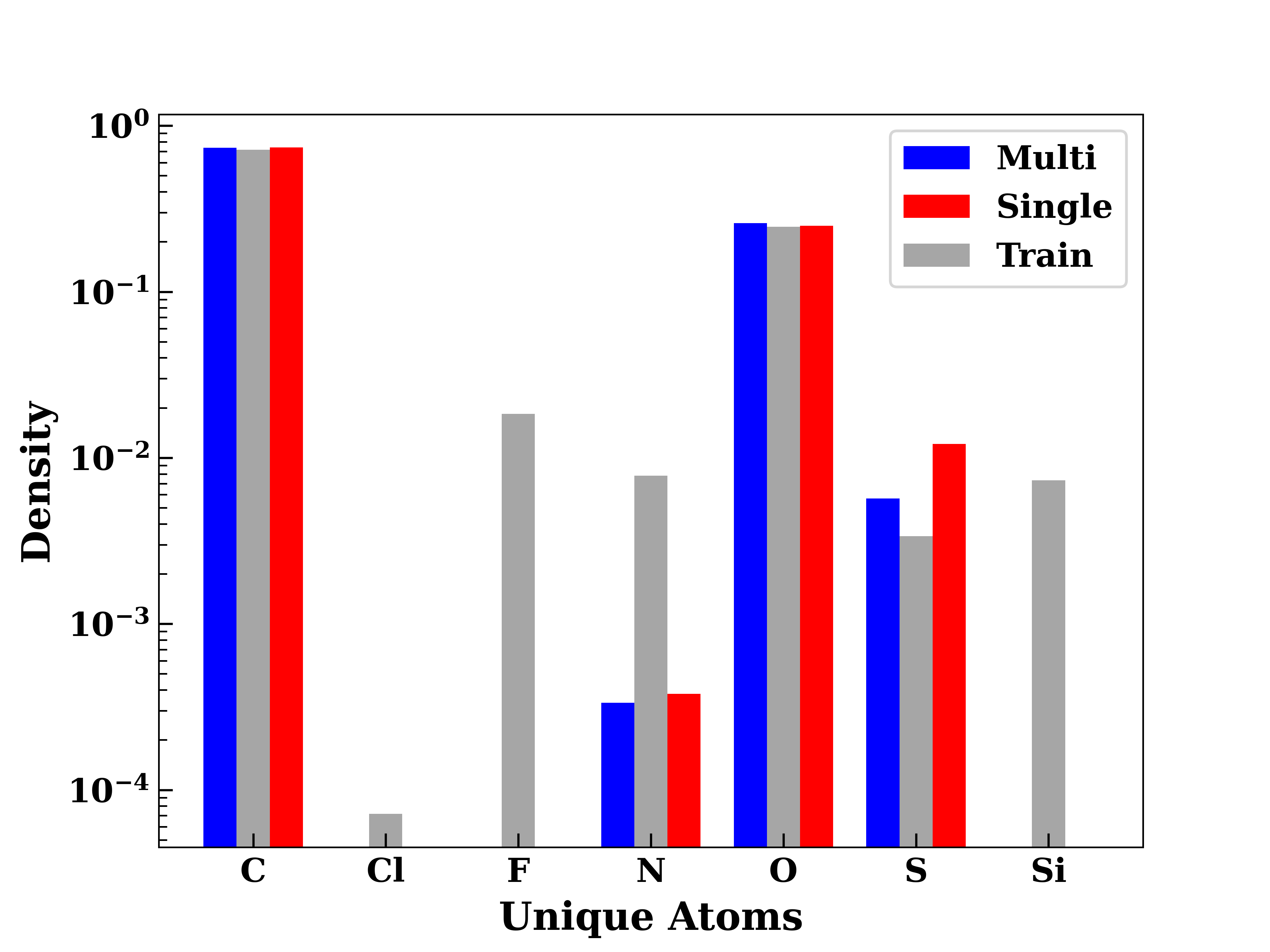}
        \caption{Transformer}
        \label{fig:atoms(b)}
    \end{subfigure}
    \caption{Distribution of heavy atoms in the generated set for different methods with red representing the distribution for molecules generated using $\text{pCMC}$ values and blue representing the molecules generated using all three property values, compared to the training set shown in grey. }
    \label{fig:atoms}
\end{figure}

\begin{figure}[t]
    \centering
    \begin{subfigure}[b]{0.45\textwidth}
        \includegraphics[width=\textwidth]{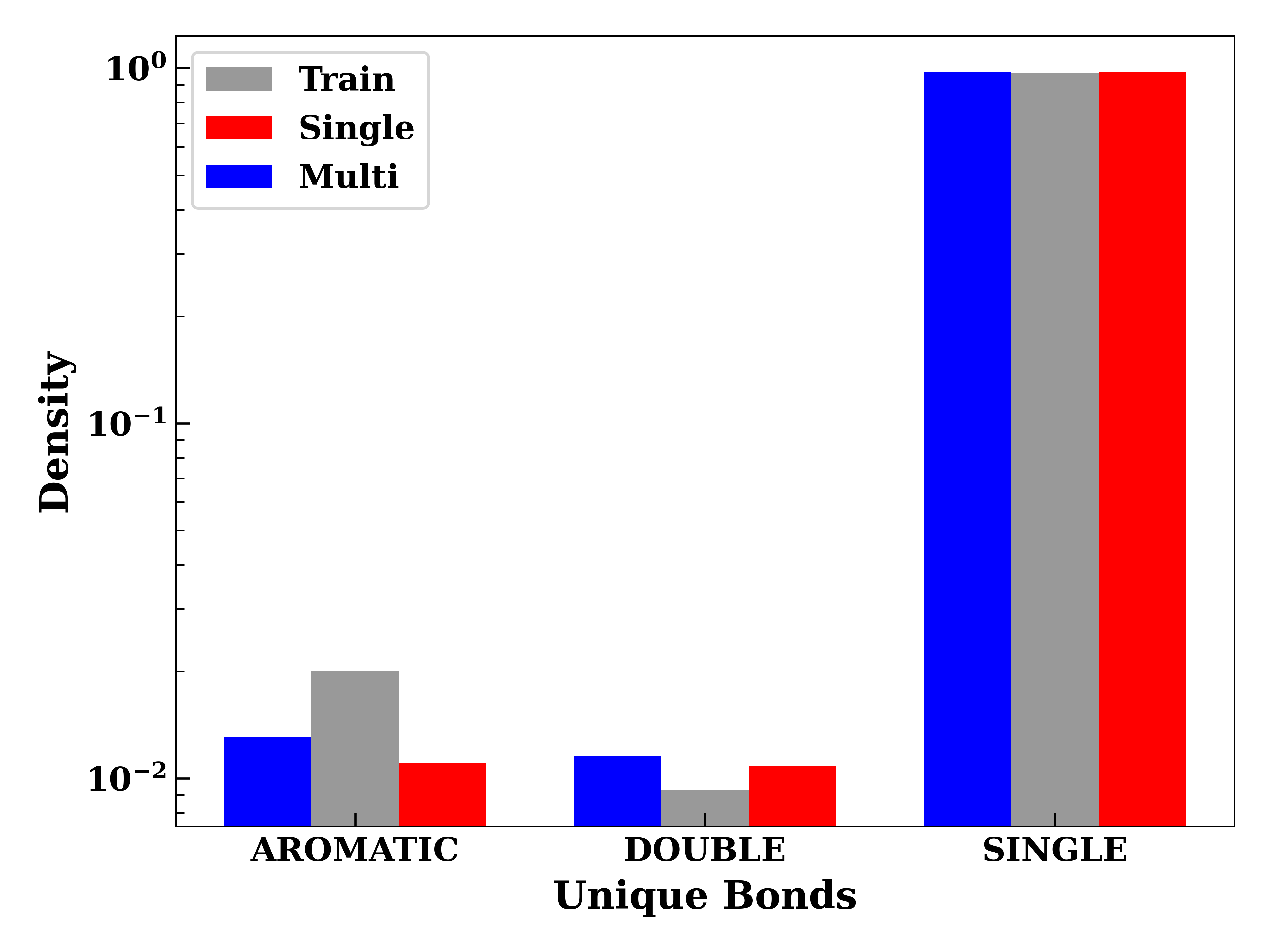}
        \caption{Diffusion}
        \label{fig:bonds(a)}
    \end{subfigure}
    \hfill
    \begin{subfigure}[b]{0.45\textwidth}
        \includegraphics[width=\textwidth]{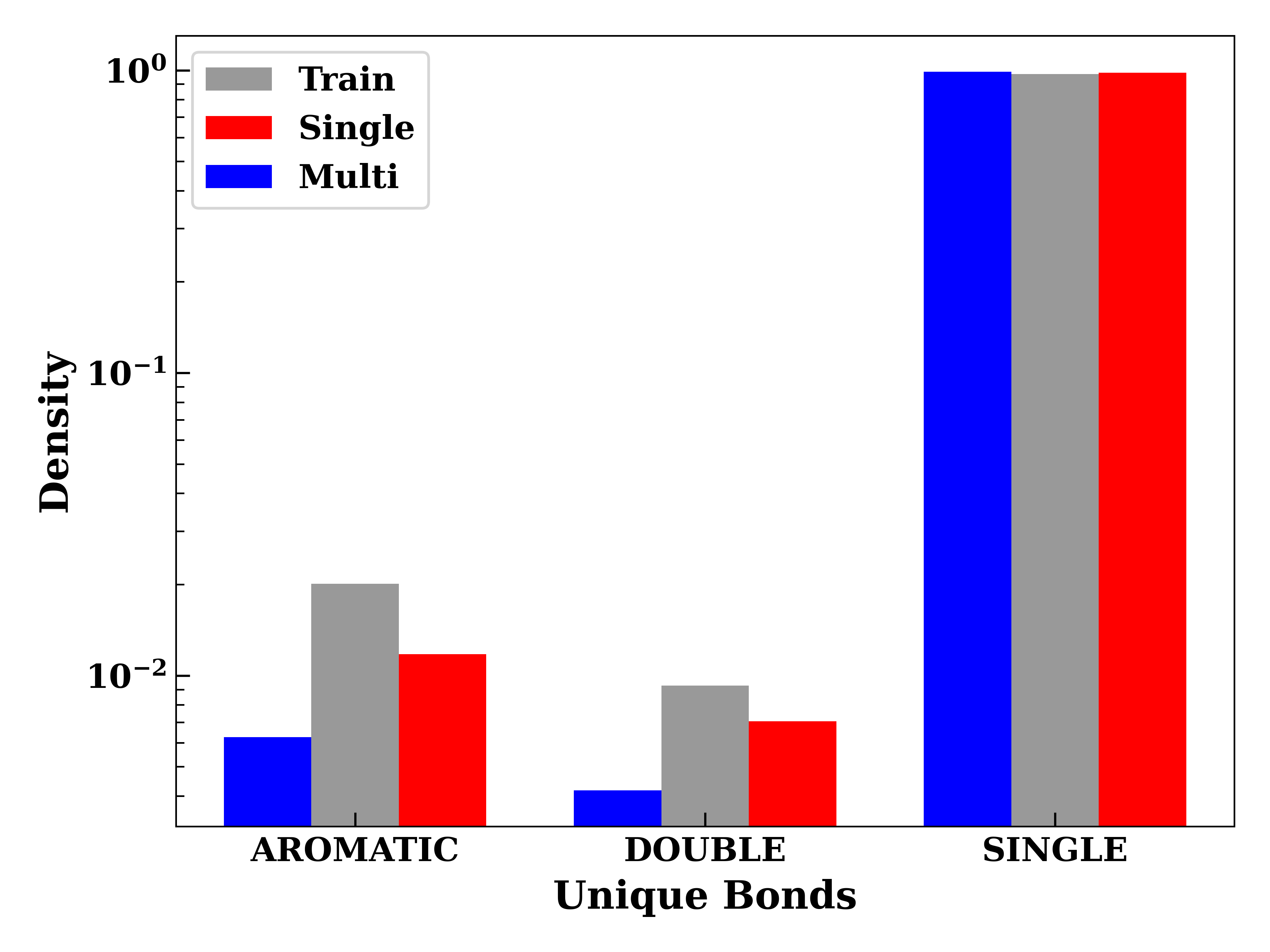}
        \caption{Transformer}
        \label{fig:bonds(b)}
    \end{subfigure}
    \caption{Distribution of unique bonds in the generated set for different methods with red representing the distribution for molecules generated using $\text{pCMC}$ values and blue representing the molecules generated using all three property values, compared to the training set shown in grey.}
    \label{fig:bonds}
\end{figure}

Figures~\ref{fig:atoms}\&\ref{fig:bonds}, also show negligible difference in the distribution of heavy atoms and bonds when comparing single and multi methods. It must be noted that for the test set in transformers, matched molecular pairs having chlorine, silicon and fluorine atoms could not be constructed.\\
The above results suggest that diffusion based inverse design models are capable sampling from a more diverse chemical space while trying to satisfy the chemical constraints, while transformer based molecule optimization methods are likely learning to sample from a tight local optimum, in the property conditioned space rather than exploring the the full chemical manifold. This explains why the generated molecules in the transformers having higher similarity scores, lower diversity scores, and don't cover the entire training space as observed in Figure~\ref{fig:tsne(b)}, despite being more efficient than diffusion models to satisfy the property constraints. In the multi-property setting, the diffusion model tends to generate compromise solutions rather than molecules that satisfy all property constraints simultaneously. This is reflected in the higher pCMC MAE of the multi-conditioned model, suggesting that enforcing several conditions at once reduces the accuracy with which each individual property is matched, while enabling the generator to explore a broader region of the training space. Such results are not observed for SMILES-based transformers, likely because adding more property conditions primarily increases the model’s context length rather than forcing joint satisfaction of all constraints. While the multi-property transformer exhibits a higher pCMC MAE, its distance scores remain close to the single-property case. This indicates that the transformer, functioning as an optimization model, remains close to the input molecule’s structure, whereas the diffusion model—operating in an inverse design setting—can move more freely through chemical space.

\subsection{Screening}
\begin{figure}[h]
    \centering
    \begin{subfigure}[b]{0.45\textwidth}
        \includegraphics[width=\textwidth]{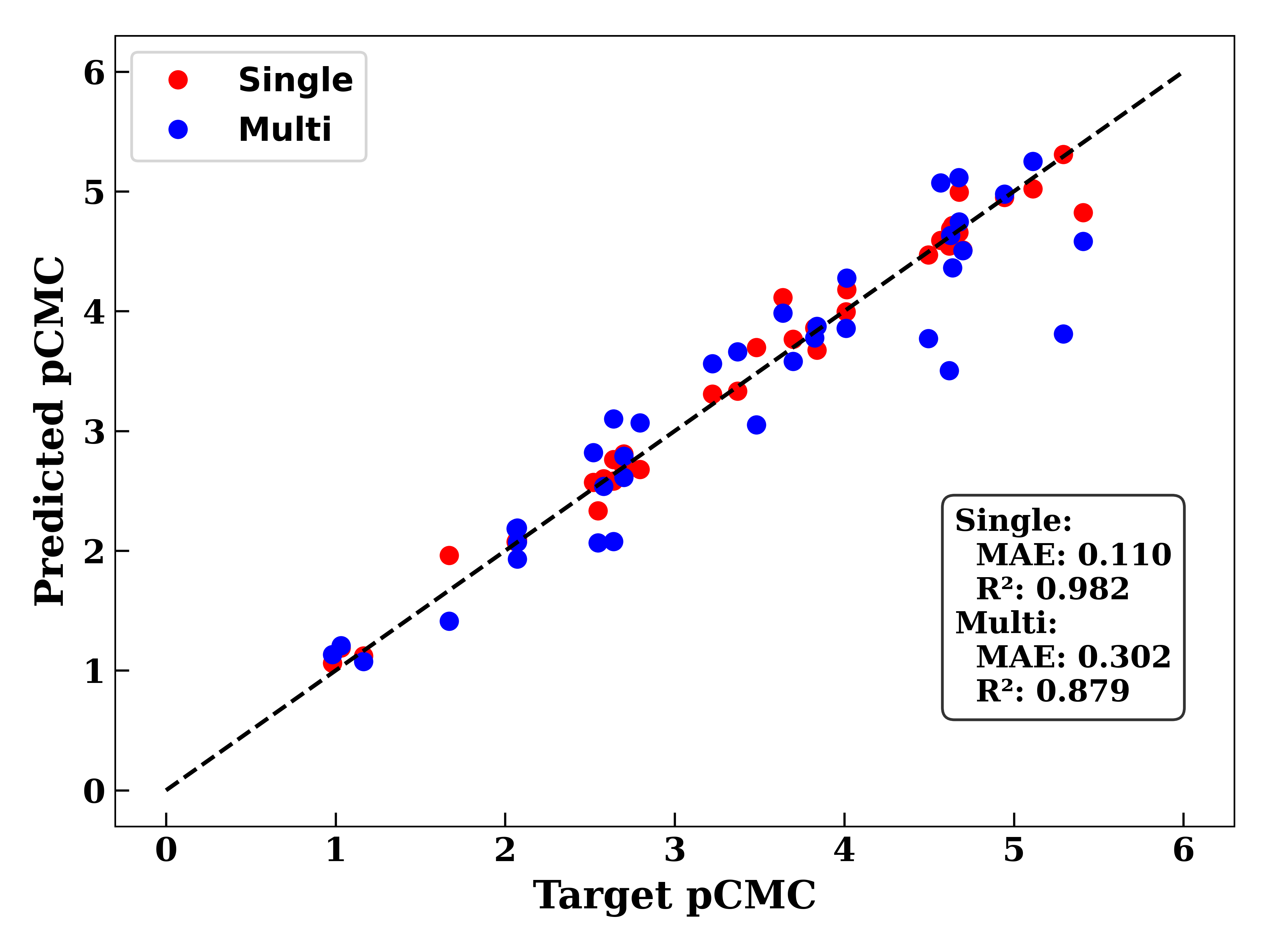}
        \caption{Diffusion}
        \label{fig:pcmc(a)}
    \end{subfigure}
    \hfill
    \begin{subfigure}[b]{0.45\textwidth}
        \includegraphics[width=\textwidth]{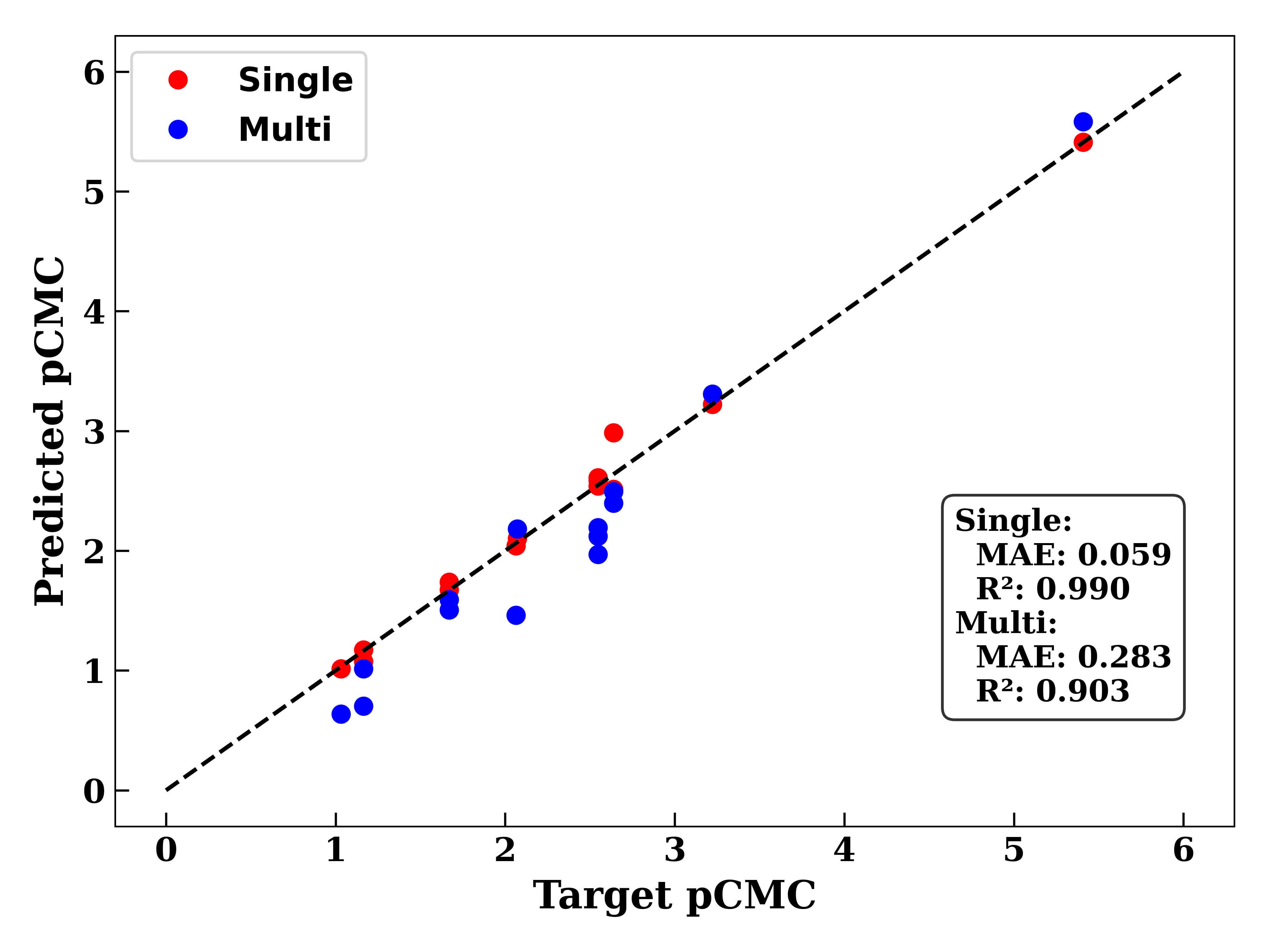}
        \caption{Transformer}
        \label{fig:pcmc(b)}
    \end{subfigure}
    \caption{Comparison of predicted vs target $\text{pCMC}$ values for final screened molecules, using trained property predictor models. Red represents the molecules generated using only $\text{pCMC}$ values while }
    \label{fig:pcmc_screened}
\end{figure}
Before validating our results through molecular dynamics simulations, we first use the trained property predictor models to screen the generated molecules based on their ability to satisfy input property conditions. In order to screen the generated molecules we first define the following metric:
\begin{equation}
    \Delta p = \frac{|p_{pred}-p_{target}|}{p_{MAE}}
    \label{eq: delta_prop}
\end{equation}

where $p$ represents any of the three property values stated earlier, $p_{pred}$ is the model prediction of that property value for a generated molecule, $p_{target}$ is the corresponding target property value, and $p_{MAE}$ represents the property predictor models' mean absolute error value for property $p$, as given in Table~\ref{tab:prop_pred}. For molecules that are generated only using $\text{pCMC}$ values, first use the $\textbf{SINGLE}$ models to predict their $\text{pCMC}$ values, and then choose the molecules with lowest $\Delta\text{pCMC}$ for each set of input conditions. For molecules generated using set of $\text{pCMC, AW\_ST\_CMC, Area\_min}$ values, we use the $\textbf{MULTI}$ models to predict all three properties, compute the metrics for each property as per Equation~\ref{eq: delta_prop} and compute the final score $d$ as: 
\begin{equation}
    d = \sqrt{\Delta\text{pCMC}^2+\Delta\text{AW\_ST\_CMC}^2+\Delta\text{Area\_min}^2}
    \label{eq: distance}
\end{equation}

Molecules which have the lowest value of $d$ are finally selected for a given set of property values. Figure~\ref{fig:pcmc_screened} shows the predicted $\text{pCMC}$ of final screened molecules compared to their target values for diffusion and transformers. The results stated in Figure~\ref{fig:pcmc_screened} again reaffirm the earlier conclusion, that inverse design models focused more on exploration, while the transformer based model tries to remain close to the input molecule's structure and learn to satisfy the input property conditions better. 

\subsection{DFT and MD Simulations }
\begin{figure}[t]
    \centering
    \begin{subfigure}[b]{0.45\textwidth}
        \centering
        \includegraphics[width=0.30\linewidth]{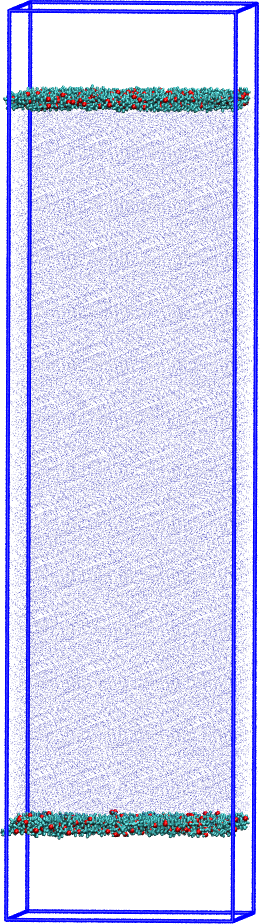}
        \caption{Initial Configuration}
    \end{subfigure}
    \hfill
    \begin{subfigure}[b]{0.45\textwidth}
        \centering
        \includegraphics[width=0.325\linewidth]{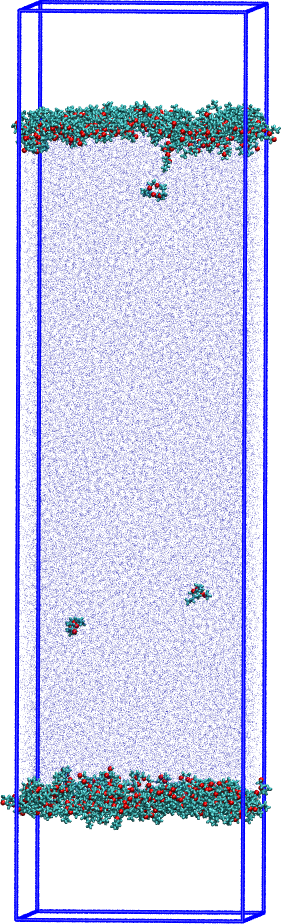}
        \caption{Final Configuration}
    \end{subfigure}
    \caption{Snapshots of configurations of generated-molecule-water system at 119 mM (CMC) concentration: The carbon and hydrogen atoms in hydrocarbon chains are rendered as cyan spheres, respectively. Orange spheres represent the oxygen atoms. Water is drawn as blue points.}
\end{figure}
In computational chemistry, optimising a molecular geometry refers to identifying the atomic arrangement that has the minimum energy. It changes the positions of the atoms and calculates the forces acting on them, gradually shifting the structure toward a point where the forces become very small. When the forces are close to zero, the molecule is considered to have reached a minimum on the potential energy surface, which represents a stable configuration that may be either a local minimum or the global minimum.

DFT calculations (structure relaxations) were carried out for all the generated molecules and the results indicates that all are structurally stable (negative energies and no imaginary frequency)hence validating the feasibility of their proposed configurations (see Table S*). \par   
The surface tension, $\gamma$ (bar nm) of the surfactant can be obtained using the following thermodynamic expression :

\begin{equation}
    \gamma = \frac{L_{z}}{2}(\bar{P}_{zz} - \frac{\bar{P}_{xx}+\bar{P}_{yy}}{2})
    \label{eq: surf_ten}
\end{equation}

where $\bar{P}_{xx}$ and $\bar{P}_{yy}$ are tangential components and $\bar{P_{zz}}$ is the normal component of the time averaged pressure tensor.

\begin{figure}[t]
    \centering
    \captionsetup{font=footnotesize}

    \begin{subfigure}[b]{0.24\textwidth}
    \begin{tikzpicture}
        \node[anchor=south west, inner sep=0] (img)
            {\includegraphics[width=\textwidth]{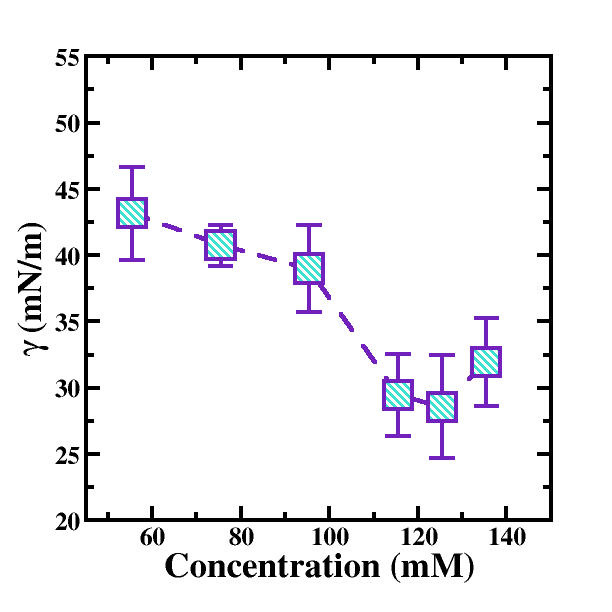}};
        \node[anchor=north east, text=black, font=\bfseries\scriptsize, yshift=-10pt, xshift=-10pt]
            at (img.north east) {(a)};
    \end{tikzpicture}
    \label{fig:img1}
    \end{subfigure}
    \hfill
    \begin{subfigure}[b]{0.24\textwidth}
    \begin{tikzpicture}
        \node[anchor=south west, inner sep=0] (img)
            {\includegraphics[width=\textwidth]{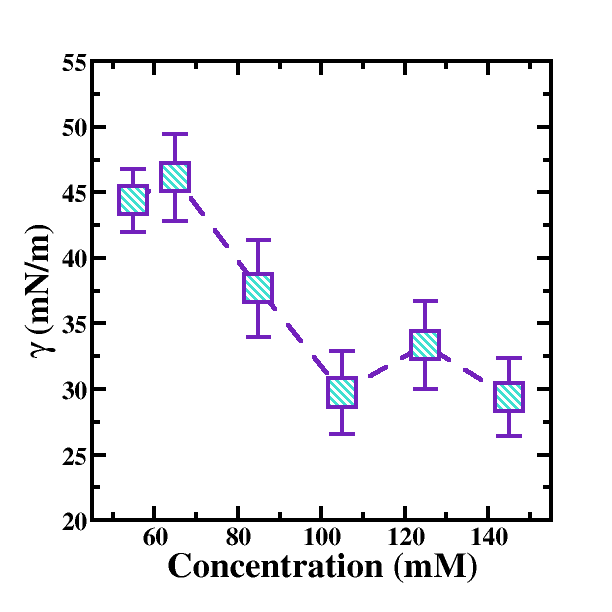}};
        \node[anchor=north east, text=black, font=\bfseries\scriptsize, yshift=-10pt, xshift=-10pt]
            at (img.north east) {(b)};
    \end{tikzpicture}
    \label{fig:img2}
    \end{subfigure}
    \hfill
    \begin{subfigure}[b]{0.24\textwidth}
    \begin{tikzpicture}
        \node[anchor=south west, inner sep=0] (img)
            {\includegraphics[width=\textwidth]{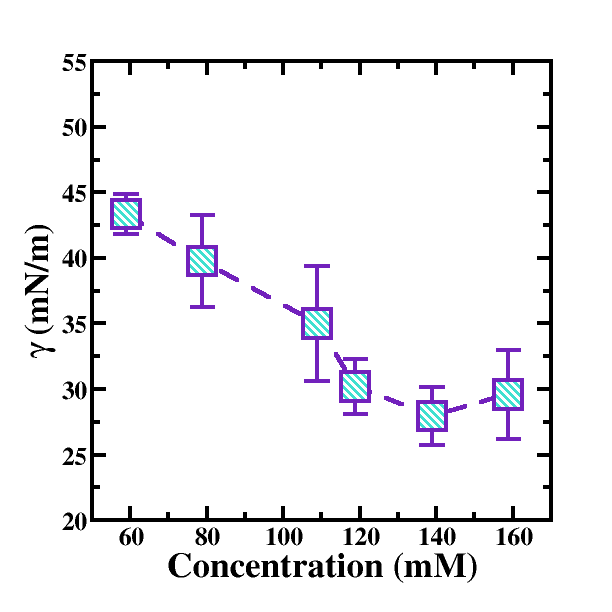}}; 
        \node[anchor=north east, text=black, font=\bfseries\scriptsize, yshift=-10pt, xshift=-10pt]
            at (img.north east) {(c)};
    \end{tikzpicture}
    \label{fig:img3}
    \end{subfigure}
    \hfill
    \begin{subfigure}[b]{0.24\textwidth}
    \begin{tikzpicture}
        \node[anchor=south west, inner sep=0] (img)
            {\includegraphics[width=\textwidth]{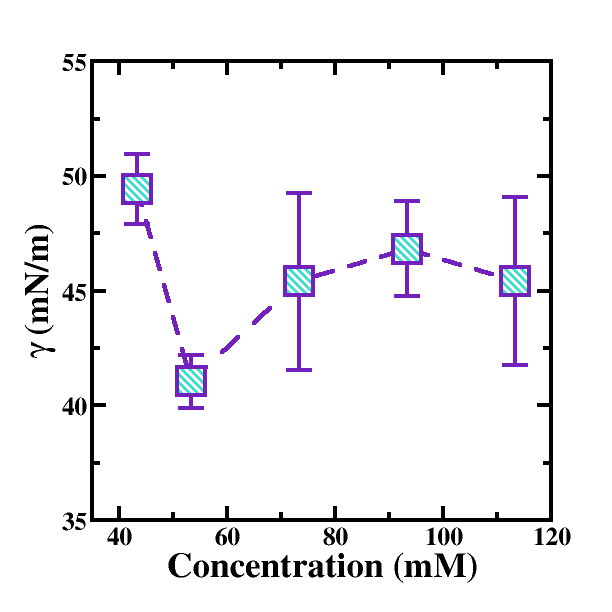}};
        \node[anchor=north east, text=black, font=\bfseries\scriptsize, yshift=-10pt, xshift=-10pt]
            at (img.north east) {(d)};
    \end{tikzpicture}
    \label{fig:img4}
    \end{subfigure}

    \vspace{1em}

    \begin{subfigure}[b]{0.24\textwidth}
    \begin{tikzpicture}
        \node[anchor=south west, inner sep=0] (img)
            {\includegraphics[width=\textwidth]{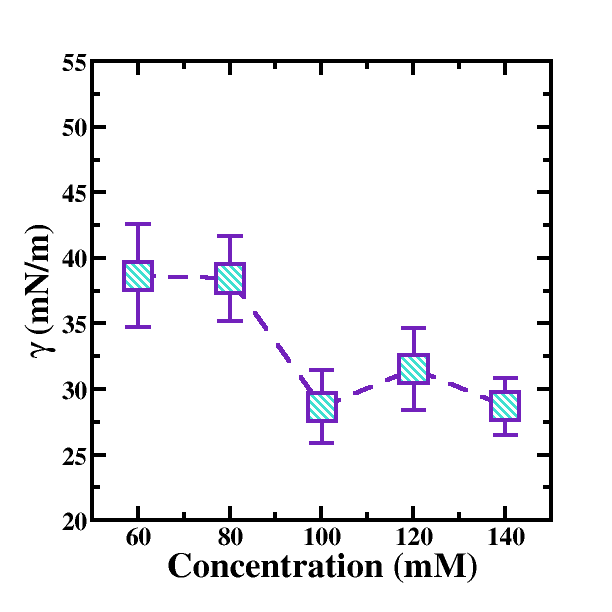}};
        \node[anchor=north east, text=black, font=\bfseries\scriptsize, yshift=-10pt, xshift=-10pt]
            at (img.north east) {(e)};
    \end{tikzpicture}
    \label{fig:img5}
    \end{subfigure}
    \hfill
    \begin{subfigure}[b]{0.24\textwidth}
    \begin{tikzpicture}
        \node[anchor=south west, inner sep=0] (img)
            {\includegraphics[width=\textwidth]{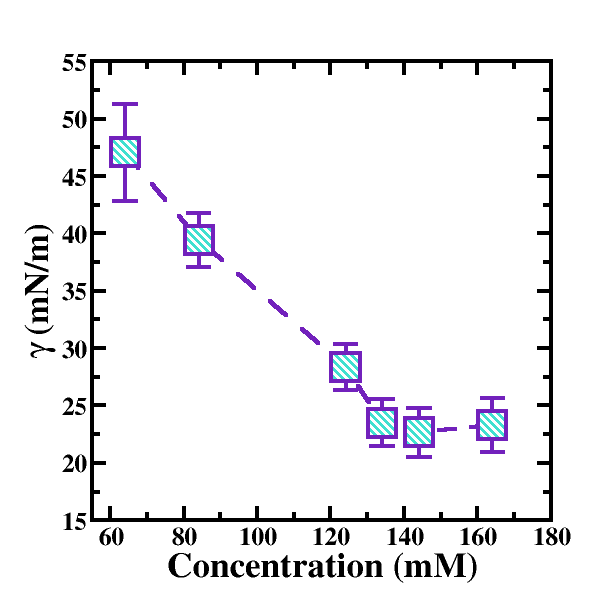}};
        \node[anchor=north east, text=black, font=\bfseries\scriptsize, yshift=-10pt, xshift=-10pt]
            at (img.north east) {(f)};
    \end{tikzpicture}
    \label{fig:img6}
    \end{subfigure}
    \hfill
    \begin{subfigure}[b]{0.24\textwidth}
    \begin{tikzpicture}
        \node[anchor=south west, inner sep=0] (img)
            {\includegraphics[width=\textwidth]{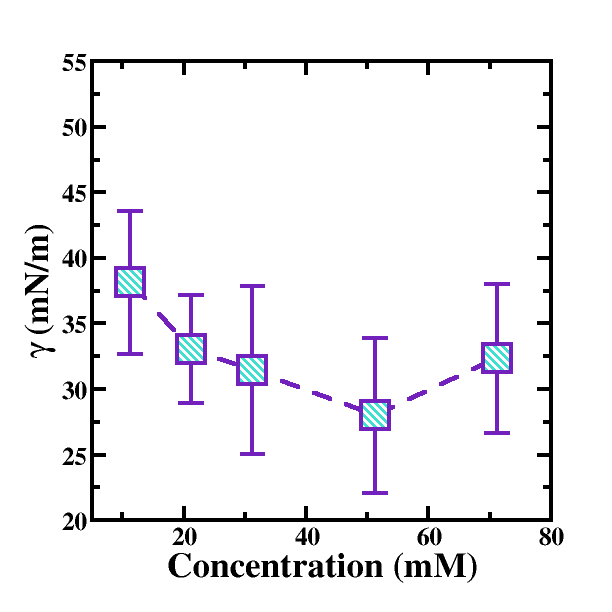}};
        \node[anchor=north east, text=black, font=\bfseries\scriptsize, yshift=-10pt, xshift=-10pt]
            at (img.north east) {(g)};
    \end{tikzpicture}
    \label{fig:img7}
    \end{subfigure}
    \hfill
    \begin{subfigure}[b]{0.24\textwidth}
    \begin{tikzpicture}
        \node[anchor=south west, inner sep=0] (img)
            {\includegraphics[width=\textwidth]{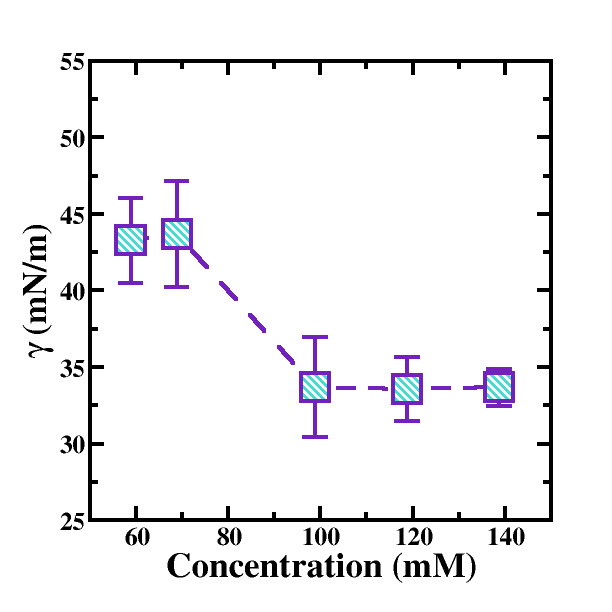}};
        \node[anchor=north east, text=black, font=\bfseries\scriptsize, yshift=-10pt, xshift=-10pt]
            at (img.north east) {(h)};
    \end{tikzpicture}
    \label{fig:img8}
    \end{subfigure}

    \caption{Equilibrium surface tension as a function of concentration for 8 screened molecules}
    \label{fig:md_vs_pred_8}
\end{figure}
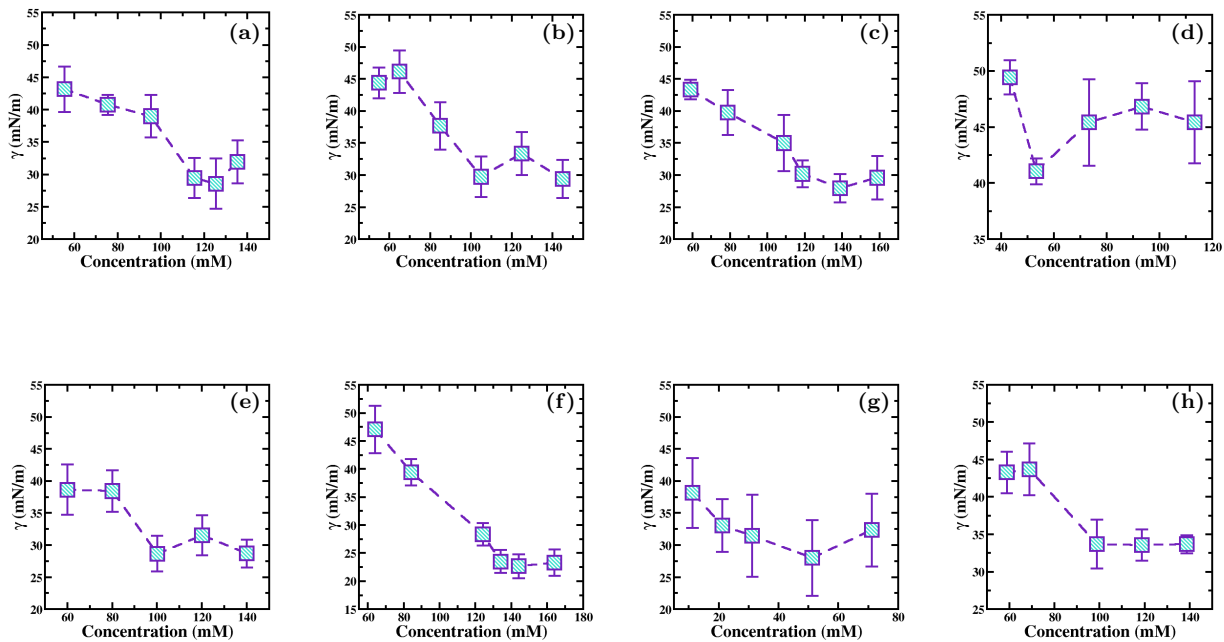

Figure~\ref{fig:md_vs_pred_8} shows the variation in the equilibrium surface tension as a function of the surfactant concentration for 8 final screened molecules which are shown in Table~\ref{tab: md_results}. Due to system constraints, the selected molecules had pCMC values in the range of approximately 0.9–1.5. The equilibrium surface tension of all of the surfactants decreases and plateaus around the critical micellar concentration as the surface is now saturated with surfactant molecules.

\begin{table}[h!]
\centering
\setlength{\tabcolsep}{10pt} 
\renewcommand{\arraystretch}{1.4} 

\scalebox{0.8}{
\begin{tabular}{>{\centering\arraybackslash}m{2.2cm}|
                >{\centering\arraybackslash}m{3.8cm}|
                >{\centering\arraybackslash}m{2.2cm}|
                >{\centering\arraybackslash}m{2.2cm}|
                >{\centering\arraybackslash}m{2.8cm}|
                >{\centering\arraybackslash}m{2.8cm}}
\textbf{Method} & \textbf{Molecule} & \textbf{Predicted pCMC} & \textbf{Simulated pCMC} & \textbf{Predicted AW\_ST\_CMC} & \textbf{Simulated AW\_ST\_CMC} \\ \hline

\multirow{2}{*}{\textbf{\begin{tabular}[c]{@{}c@{}}Diffusion\\ Single\end{tabular}}}
& \raisebox{-0.4\width}{\includegraphics[width=4.0cm]{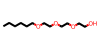}} & 1.122 & 0.9375 & 30.3945 & 29.4423 \\ \cline{2-6}
& \raisebox{-0.4\width}{\includegraphics[width=4.0cm]{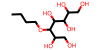}} & 1.1876 & 0.9791 & 30.4735 & 29.7395 \\ \hline

\multirow{2}{*}{\textbf{\begin{tabular}[c]{@{}c@{}}Diffusion\\ Multi\end{tabular}}}
& \raisebox{-0.4\width}{\includegraphics[width=4.0cm]{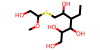}} & 0.9248 & 0.9248 & 29.0643 & 30.1967 \\ \cline{2-6}
& \raisebox{-0.4\width}{\includegraphics[width=4.0cm]{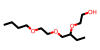}} & 1.1343 & 1.1343 & 29.2732 & 45.4178 \\ \hline

\multirow{2}{*}{\textbf{\begin{tabular}[c]{@{}c@{}}Transformer\\ Single\end{tabular}}}
& \raisebox{-0.4\width}{\includegraphics[width=4.0cm]{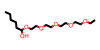}} & 0.9996 & 0.9996 & 30.6493 & 28.6462 \\ \cline{2-6}
& \raisebox{-0.4\height}{\includegraphics[width=4.0cm]{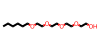}} & 1.0753 & 0.8729 & 31.4097 & 23.5044 \\ \hline

\multirow{2}{*}{\textbf{\begin{tabular}[c]{@{}c@{}}Transformer\\ Multi\end{tabular}}}
& \raisebox{-0.4\width}{\includegraphics[width=4.0cm]{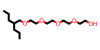}} &1.5052 & 1.0052 & 30.2822 & 31.4467 \\ \cline{2-6}
& \raisebox{-0.4\width}{\includegraphics[width=4.0cm]{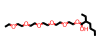}} & 1.0052 & 1.0052 & 33.7755 & 33.6901 \\
\end{tabular}}
\caption{Results of Molecular Dynamics Simulations}
\label{tab: md_results}
\end{table}

The comparison of the predicted and simulated values of surface tension exhibited good agreement. This implies that the predictive model performs quite well in the selection of surfactants prior to running detailed simulations.The estimated surface tension deviates from the experimental values in some cases as seen in Table~\ref{tab: md_results}, which may be attributed to
(i) system size effects, (ii) the shortcomings of the model and (iii) absence of the long range correction to the surface tension.\cite{tiwari2022effect}


In summary, the MD simulation results again confirm the reliability of the predictive framework developed earlier, showing explicit connections between molecular structure and interfacial behavior. This method thus allows for an efficient way of accelerating the discovery and design of new surfactant molecules by coupling machine learning-based screening with MD validation.

\section{Conclusions and Future Work}

We evaluate the conditional generative capabilities of two distinct generative frameworks—diffusion (inverse design) and transformers (optimization-based)—for generating non-ionic surfactants under single- and multi-property conditions. We find that diffusion produces a more diverse set of candidates, with diversity scores of $0.836$ (single-property) and $0.840$ (multi-property), compared to transformers with scores of $0.661$ and $0.640$. This reflects the diffusion model’s ability to explore broader chemical regions rather than remaining anchored to an input structure.\\
In contrast, using trained property predictor models we see that transformers achieve more accurate property satisfaction: the average pCMC MAE for transformers is $0.482$ in the single-property setting and $0.695$ in the multi-property setting, compared to diffusion’s $0.975$ and $1.306$. This indicates that transformers better match target constraints but at the cost of reduced chemical diversity.\\
We further validate selected generated molecules using molecular dynamics simulations and find that predicted properties agree reasonably with simulated values, confirming that both models can generate chemically valid molecules that satisfy input conditions.\\
Across both model classes, we observe that multi-property conditioning increases the average pCMC error (e.g., diffusion: $0.975\to1.306$; transformer: $0.482\to0.695$), suggesting that adding constraints makes it harder for the models to satisfy each target simultaneously. Despite this, diversity remains stable across single- and multi-property modes for both architectures, even allows multi-property diffusion models generator to explore broader regions of the training space as seen by lower distance scores ($6.457\to 5.260$).\\
Overall, diffusion offers broader exploration and higher structural diversity, whereas transformers offer tighter property alignment but more limited search around the input. Extending these results to both ionic and non-ionic surfactants and incorporating a richer property set remains an area for future work.

 \paragraph*{}

\subsection{Author information}

\subsubsection{Corresponding Authors}

\textbf{Jayant K. Singh} --- Chemical Engineering, Indian Institute of Technology Kanpur,
Kanpur 208016, India; Prescience Insilico Private Limited, Bangalore 560049, India;
\href{https://orcid.org/0000-0001-8056-2115}{orcid.org/0000-0001-8056-2115};
Phone: +91-512-259 6141; Email: jayantks@iitk.ac.in; Fax: +91-512-259 0104.\\

\noindent
\textbf{Sudip Roy} --- Prescience Insilico Private Limited, Bangalore 560049, India

\subsubsection{Authors}

\textbf{Sayeedul I. Sheikh} --- Chemical Engineering, Indian Institute of Technology Kanpur,
Kanpur 208016, India; Email: sayeedul21@iitk.ac.in.\\
\noindent
\textbf{V. Subhasree Navya} --- Chemical Engineering, Indian Institute of Technology Kanpur,
Kanpur 208016, India; Email: vsubhasree24@iitk.ac.in.

\noindent
\textbf{Riya Sharma} --- Chemical Engineering, Indian Institute of Technology Kanpur,
Kanpur 208016, India\\
\subsection{Author’s contributions}

S.I.S.,V.S.N. and R.S. contributed equally to the investigation, validation, organization of the data, and visualization of the work. V.S.N., R.S., and J.K.S. conceptualized the project. S.I.S. and V.S.N. formulated the methodology and performed the validation, while R.S. carried out the formal analysis of the work. S.I.S., V.S.N. and R.S. contributed equally to writing the paper. R.S., J.K.S., and S.R. coordinated the project, and J.K.S. organized the resources and funding.

\section{Acknowledgement}
S.I.S., V.S.N. and R.S. would like to acknowledge the Param HPC resources at IIT Kanpur. J.K.S. would also like to acknowledge Center of Excellence for Speciality Chemicals (MCF/CHE/2024530). R.S. would also like to thank the Ministry of Education, for providing financial assistance through the Prime Minister Research Fellowship (2303414).

\suppinfo
Simulation methodology and details are provided in Table S1 of the Supporting Information. Codes used in the paper is available at \url{https://github.com/sayeed02021/digital-surfactant}
\bibliography{acs-references}
\end{document}